\documentclass[12pt,preprint]{emulateapj}
\newcommand\kms{\mbox{km~s$^{-1}$}}%
\newcommand\rotate{\@pt@rottrue}%
\shorttitle{SAGITTARIUS B2 MAIN}
\shortauthors{Zhao \& Wright}
\begin{document}
%\twocolumn[
\title{SAGITTARIUS B2 MAIN: A CLUSTER OF ULTRA-COMPACT HII REGIONS \& MASSIVE PROTOSTELLAR CORES}
\author{Jun-Hui Zhao\altaffilmark{1} and M. C. H. Wright\altaffilmark{2}}
%%%%%%%%%%%%%%%%%%%%%%%%%%%%%%%%%%%%%%%%%%%%%%%%%%%%%%%%%%%%%%%%%%%
%%%%%%%%%%%%%%%%%%%%%%%%%%%%%%%%%%%%%%%%%%%%%%%%%%%%%%%%%%%%%%%%%%%
\begin{abstract}
The ionized core in the Sgr B2 Main star-forming region was imaged
using the Submillimeter Array archival data observed for the H26$\alpha$ 
line and continuum emission at 0.86 millimeter with an angular resolution 
0.3\arcsec. Eight hyper-compact H26$\alpha$ emission sources were 
detected with a typical size in the range of 1.6--20$\times10^2$ AU and 
electron density of 0.3--3$\times10^7$ cm$^{-3}$, corresponding to the 
emission measure 0.4--8.4$\times10^{10}$ cm$^{-6}$ pc. 
The H26$\alpha$ line fluxes from the eight hyper-compact HII sources 
imply that the ionization for each of the sources must be powered by 
a Lyman continuum flux from an O star or a cluster of B stars. 
The most luminous H26$\alpha$ source among the eight detected requires an O6 
star that appears to be embedded in the ultra-compact HII region F3.
In addition, $\sim$ 23 compact continuum emission sources
were  also detected within the central 5\arcsec$\times$3\arcsec\,($\sim0.2$ pc) 
region. In the assumption of a power-law distribution for the dust temperature,
with the observed brightness temperature of the dust emission we determined 
the physical properties of the submillimeter emission sources showing that 
the molecular densities are in the range of 1--10$\times10^8$ cm$^{-3}$, 
surface densities between 13 to 150 $g$ cm$^{-2}$, and total gas masses 
in the range from 5 to  $\gtrsim$ 200 $M_\odot$ which are
1 or 2 orders of magnitude greater than the corresponding values
of the Bonnor-Ebert mass. With a mean free-fall time scale of 2$\times10^3$ y,
each of the massive protostellar cores are undergoing gravitational collapse 
to form new massive stars in the Sgr B2 Main core.  
\end{abstract}
%%%%%%%%%%%%%%%%%%%%%%%%%%%%%%%%%%%%%%%%%%%%%%%%%%%%%%%%%%%%%%%%%%%
%%%%%%%%%%%%%%%%%%%%%%%%%%%%%%%%%%%%%%%%%%%%%%%%%%%%%%%%%%%%%%%%%%%
  %]
\altaffiltext{1}{Harvard-Smithsonian Center for Astrophysics, 60
  Garden Street, Cambridge, MA 02138, USA; jzhao @ cfa.harvard.edu}
\altaffiltext{2}{Department of Astronomy, University of California, Berkeley, Berkeley, CA 94720, USA}

%%%%%%%%%%%%%%%%%%%%%%%%%%%%%%%%%%%%%%%%%%%%%%%%%%%%%%%%%%%%%%%%%%%
%%%%%%%%%%%%%%%%%%%%%%%%%%%%%%%%%%%%%%%%%%%%%%%%%%%%%%%%%%%%%%%%%%%
\section{INTRODUCTION}
The formation of massive stars is one of the challenging problems 
in astrophysics. Unlike their low-mass countparts, massive stars 
are rare, and form in relatively deeply embedded massive molecular 
clouds on a much shorter time-scale \citep{mcla97,osor99,mcke02}. High-mass 
stars are often formed in clusters\citep{mcke07}. After initial gravitational 
collapse of the natal clouds, multiple protostellar cores are 
formed through fragmentation of the gas. One of the fundamental 
theoretical questions is how the massive stars form in clusters.
According to the competitive theory  \citep{bonn04,bonn06,bonn08},
mass clumps created from natal cloud gravitational collapse contain
protostellar cores with small masses. These cores
grow by accreting matter, competing with other cores, 
and form stars with many times their original mass. On the other 
hand, the direct gravitational collapse theory \citep{krum05}
suggests that the protostellar cores created from fragmentation 
of the initial cloud gravitational collapse have sufficient mass 
to form individual high-mass to low-mass star systems in the subsequent 
collapse. Accretion from the parental cloud continues but does 
not substantially change their mass.

Strong radiation pressure from a newly formed massive star
might halt infall, limiting the mass of  stars that can form
\citep{kahn74,wolf87,lars71}. Recent theoretical investigations
suggest that the radiation pressure limit might be overcome in actual
cases with complex, non-spherical infall geometries  or  high ram pressures
in rotating disks \citep{mcke02,krum09,naka89}, or massive stars may form
from stellar merging in a dense cluster \citep{bonn98,bonn02}. 
Numerical simulations suggest that graviational instabilities cause the disk
to fragment and form a massive companion to the primary \citep{krum09},
and  consequently  form binaries in a dense stellar system
\citep{bonn02}. Furthermore,  radiation feedback from  massive stars
in a cluster affects its subsequent fragmentation and consequently plays
an important role in determining the stellar initial
mass function (IMF) in a cluster \citep{krum10}.

In the past decade, good 
progress has been made in understanding massive young stellar clusters 
in the Galaxy. A recent review \citep{fige08} shows ten known Galactic 
clusters with masses $\le10^4\, M_\odot$ with ages of a few million years. 
The Arches cluster in the Galactic center is the densest young cluster in 
the Galaxy and contains a large collection of massive stars  
\citep{fige05}. Three out of
ten (Quintuplet, Arches and Center) are located within 
the central 50 pc of the Galaxy, suggesting
that the Galactic central region appears to prefer forming massive stars. 

Located at a distance of 7.8 kpc \citep{reid09} towards the Galactic center, 
Sgr B2, a giant molecular cloud(GMC) with a mass of 6$\times10^6\, M_\odot$ 
\citep{gold90} is one of the most active star-forming region in the Galaxy, 
radiating 
a total luminosity of 1$\times10^7\, L_\odot$ \citep{gold90,gold92}. 
As the most luminous core among the several cores in this GMC, Sgr B2 Main 
is associated with numerous ultra-compact (UC) HII regions, suggesting 
the presence of a tight cluster of OB stars \citep{gaum90,gaum95,dpre98}. 
The observed molecular outflows and infalls suggest that on-going star 
formation activities are taking place \citep{qin08,rolf10}. Sgr B2 Main 
appears to be in a very young phase of forming a massive stellar cluster 
from the dense molecular core.  
   
High-resolution observations using the Submillimeter Array
(SMA)\footnote{The Submillimeter Array is a joint project 
between the Smithsonian Astrophysical Observatory and the 
Academia Sinica Institute of Astronomy and Astrophysics 
and is funded by the Smithsonian Institution and the Academia 
Sinica.} at submillimeter wavelengths can explore the 
detailed structure of  the massive star forming 
core, providing useful clues on how massive stars 
form in a cluster. 

The rest of this paper is organized as follows. In Section 2, 
we discuss the  reduction and imaging process of the SMA archival 
data of Sgr B2 Main observed in 2007. Section 3 shows the 
results from the high-resolution observations of the H26$\alpha$ 
line and continuum at 0.86 mm. Section 4 presents a model for the 
H26$\alpha$ sources. In Section 5, the properties of the bright, 
compact dust emission clumps are determined and discussed. 
In Section 6, we discuss the kinematics in terms of
ionized outflows/expansions and rotating disks. The 
early phase of massive star formation, and origin of the massive
protostellar clumps are also discussed. In Section 7,  we summarize
the conclusions.
%--------------line images-----------------------
\begin{figure*}[t]
\centering
\includegraphics[width=130mm,angle=-90]{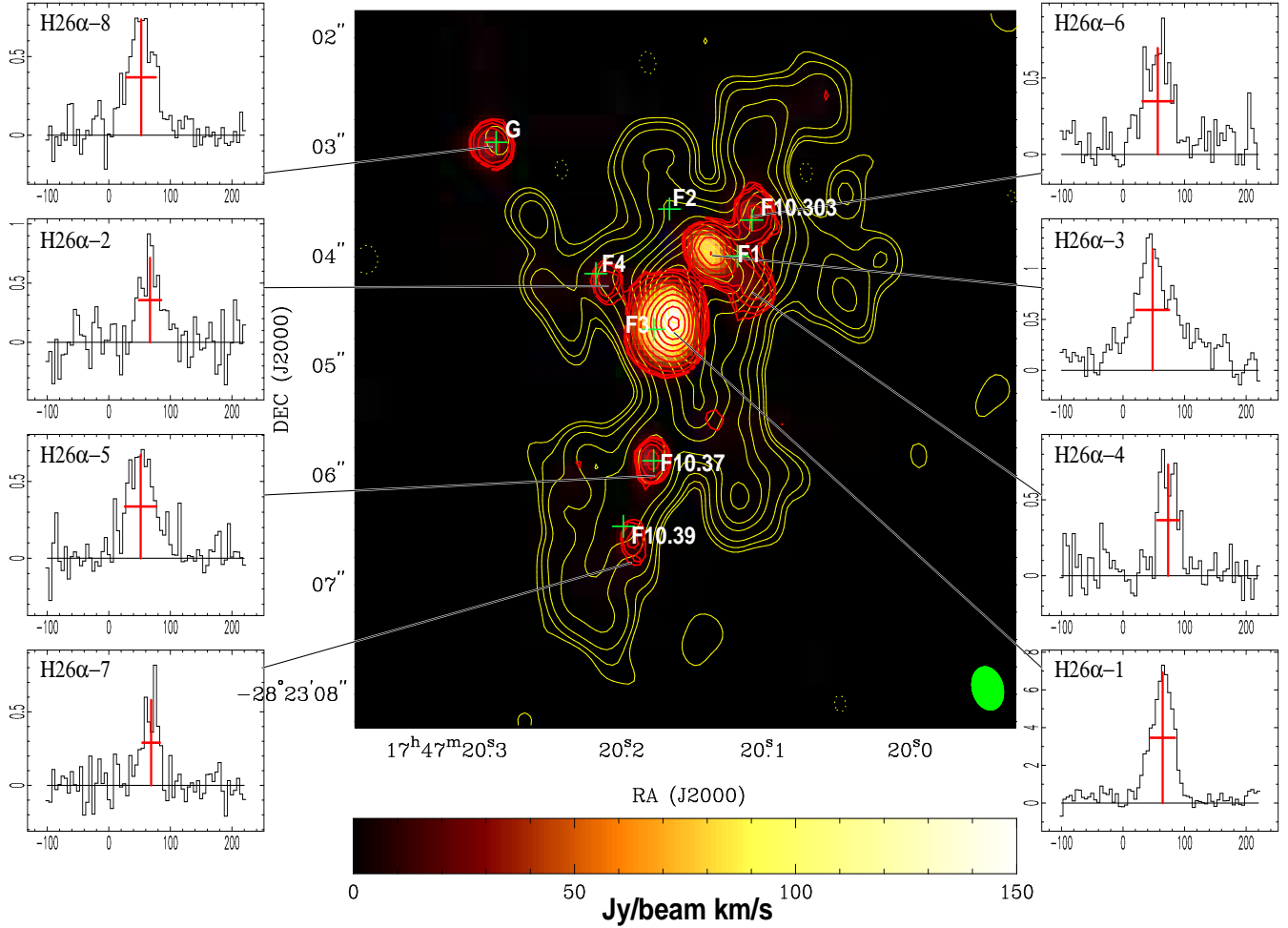}
\caption{
{\bf a:} SMA image of H26$\alpha$ line emission at the rest frequency 
of 353.623 GHz, integrated from the velocity range of $-100$ 
to $125$~km~s$^{-1}$ (red contours and color) overlaid on the 
continuum image at  0.86\,mm (yellow contours). The red contours 
are $F_{\rm P}({\rm H}26\alpha) \times $ (4, 5, 7, 11, 17, 25, 35, 
47, 61, 77 and 95\%), and the yellow contours are $S_{\rm P}(0.86 mm) 
\times $ (-2, 2, 3, 5, 10, 15, 25, 35, 45, 55, 65, 75, 85, and 95\%),
where the peak line flux $F_{\rm P}({\rm H}26\alpha)$ and peak 
continuum flux density $S_{\rm P}(0.86 mm)$ are 193 
Jy\,beam$^{-1}$\,km\,s$^{-1}$ and 2.88 Jy\,beam$^{-1}$, respectively. 
The FWHM beam is 0.40\arcsec$\times$0.28\arcsec\,(15\arcdeg) denoted 
at bottom-right. The color wedge scales the H26$\alpha$ line intensity 
in the units of Jy\,beam$^{-1}$\,km\,s$^{-1}$ of the UC HII
regions. The spectra from the eight brightest H26$\alpha$ sources 
are also shown. The units of the vertical (line flux density) and 
horizontal (LSR velocity) axes in the spectral plots are Jy and km\,s$^{-1}$, 
respectively. The red cross in each spectral profile shows the peak 
line intensity (vertical bar) and the FWHM velocity width
of a Gaussian profile determined from least-square fitting.
}
\label{fig:H26aline}
\end{figure*}

%%%%%%%%%%%%%%%%%%%%%%%%%%%%%%%%%%%%%%%%%%%%%%%%%%%%%%%%%%%%%%%%%%%
%%%%%%%%%%%%%%%%%%%%%%%%%%%%%%%%%%%%%%%%%%%%%%%%%%%%%%%%%%%%%%%%%%%
\section{Observations and Data reduction}
The interferometer data for the H26$\alpha$ line at $\nu_0=$
353.623 GHz were acquired from the SMA archive, observed
2007 June 18,  with the ``very extended'' 
array configuration in the upper-side band (USB).  The reduction for
the line data was made in Miriad \citep{STW95} following the
reduction instructions for SMA data\footnote{\it
http://www.cfa.harvard.edu/sma/miriad}.
The bandpass calibration was made by applying a normalized average
of  all the data from the QSOs included in the observing run
(J1229+020, J1733-130, J1743-038, J1751+096 and J2015+371).
The flux density scale was determined by comparing observations
of Callisto with the SMA planetary model. An
angular size and brightness temperature of 1.5\arcsec\, and 120 K
at the observing epoch were assumed. The complex gains of the data
were primarily calibrated using the nearby QSO J1733-130
to obtain an image of the SgrB2 Main core. This image 
was used as an initial model to further correct  
residual errors in the visibilities   
using self-calibration.

To separate line and continuum emission, we
used the Miriad task UVLIN. The line visibility data set was constructed
by subtructing the continuum  which was determined from  line-free channels.
We made images of the  H26$\alpha$ line with a channel width 
3 km s$^{-1}$ using a
robustness weighting  of 2 corresponding to natural 
weighting. The FWHM beam is $0.40\arcsec\times0.28\arcsec$ 
(PA=15$\arcdeg$).  The typical rms noise in a channel image is  
0.1 Jy~beam$^{-1}$.

% next 2 paragraphs edited 07 June 11:50 am
A line-free continuum data set was also produced from the UVLIN program. 
We used both USB and LSB data to image the continuum emission with
robustness weighting parameters 2 and --2 which correspond to 
natural and uniform 
weighting respectively.  The FWHM beam with  uniform weighting
is $0.36\arcsec\times0.22\arcsec$ (PA=13$\arcdeg$). The typical rms
noise in the continuum images are  7.5 and 8.5 mJy~beam$^{-1}$ for natural
and uniform weighting, respectively.
The shortest  sampled spatial frequency in the visibility data is 
30 k$\lambda$, corresponding to an angular size $\sim$ 
6\arcsec--7$\arcsec$ for the H26$\alpha$ line and continuum emission 
structure. The compact structure ($<$3\arcsec) in the H26$\alpha$ 
line and continuum emissions from the Sgr B2 Main core    
has been adequately sampled in the SMA observations. 

We noticed a systematic offset of 0.3\arcsec\,  in the phase 
center of the SMA images by comparing the positions of the 
UC HII regions G and F10.37 determined at 22.4 GHz with the 
VLA \citep{gaum95}. The offset could reflect 
residual errors in antenna positions and complex gain calibration due to the 
large distance between the calibration QSO J1733-130 and 
SgrB2 Main. We therefore imposed a  shift of 0.3\arcsec\,
for the phase center of the SMA data to align with the VLA image. 
The positional error of a compact source with $S/N=6$ in the 
final SMA images with respect to the VLA coordinate frame 
is $\sim$0.03\arcsec (see below).

\section{Results}
\subsection{H26$\alpha$ line emission}
\subsubsection{Distribution}
Figure \ref{fig:H26aline}a shows the integrated H26$\alpha$ line
image of Sgr B2 Main made from  line emission in the 
LSR velocity range between $-$100 and +225~km~s$^{-1}$. Eight 
sources with significant H26$\alpha$ line emission ($>6\sigma$ 
in the channel image) are labelled as H26$\alpha$-n, 
where n is  numbered from 1 to 8. The source H26$\alpha$-1 
has a small angular offset from the brightest 
continuum source (F3) at 22.4 GHz in the core of Sgr B2 Main while  
H26$\alpha$-8 has a larger angular distance from 
F3 \citep{gaum95}.
The two relatively isolated bright line sources, H26$\alpha$-8 and 
H26$\alpha$-5, were used to align the 0.86 mm coordinate frame of 
our SMA images with the coordinate frame dertermined with the VLA at 
22.4 GHz. The image shows the submillimeter positions of H26$\alpha$-8 
and H26$\alpha$-5 agree with the 22.4 GHz positions of G and F10.37  
within $\sim$0.01\arcsec. 
H26$\alpha$-6 appears to coincide with F10.303.
However, H26$\alpha$-1, the brightest line source, is located 0.17\arcsec\, 
northwest of F3. H26$\alpha$-6  coincides with F10.303. Two ultra-compact 
H26$\alpha$ line sources (3 and 4) appear to be associated with F1.
The hyper-compact H26$\alpha$ sources 3 and 4  are located $\sim$0.15\arcsec\, 
northeast and southwest of F1, respectively. Also, 
H26$\alpha$-2 is located 0.2\arcsec\, southwest of F4.
No significant H26$\alpha$ emission associated with F2 has been detected. 
The positional offsets between the inner 
four H26$\alpha$ emission sources (1,2,3 and 4) and their 
corresponding 22.4 GHz counterparters (F3, F4 and F1)
appear to be signficant, reflecting the fact that the 
free-free continuum emission at 22.4 GHz
traces relatively lower-density ionized gas in the outer 
layer of the expanding HII gas or the ionized outlows,
while the H26$\alpha$ line traces the  high-density ionized 
regions, possibly associated with the ionized disks or ultra- 
or hyper-compact ionized cores where the free-free emission 
appears to be optically thick at 22.4 GHz.

\subsubsection{Line broadening}
The spectral profiles of the H26$\alpha$ line from the eight brightest 
UC HII regions in the core of Sgr B2 Main are shown in Figure \ref{fig:H26aline}.
H26$\alpha$-1, associated with F3, has a peak velocity
V$_{\rm LSR}=64\pm1$ km s$^{-1}$ with a line width 
$\Delta V_{\rm FWHM}=40\pm1$ km s$^{-1}$  compared with  
V$_{\rm LSR}=68\pm2$ km s$^{-1}$
and $\Delta V_{\rm FWHM} =63\pm5$ km s$^{-1}$ derived from the VLA
measurements of the H66$\alpha$ line \citep{dpre96}.
The peak velocity between the H26$\alpha$ and H66$\alpha$ line
profiles shows a 2$\sigma$
difference. The line width of the H66$\alpha$ is much broader
than that of H26$\alpha$.  Using canonical values of  
$T_{\rm e}=1\times10^4$ K and n$_{\rm e}=1.1\times10^6$ cm$^{-3}$ 
for the typical UC HII regions in Sgr B2 Main \citep{dpre98}, 
for the VLA observations of the continuum emission 
at 7 mm, the thermal broadening $\Delta V_{\rm th}$ contributes 
21 km s$^{-1}$ to the line widths. The pressure broadening depends 
upon the principal quantum number $N$ and electron density 
$\rm n_{\rm e}$ ($ \Delta V_{\rm P} \propto n_{\rm e}\,N^{7.4}$) \citep{brock71,brock72}. 
We find 
that the pressure broadening ($\Delta V_{\rm P}\sim0.02$ km s$^{-1}$) is 
negligible for the H26$\alpha$ line.  Thus, the non-thermal broadening $\Delta V_{\rm nth}$  
including the Doppler motions of expansion, infall, outflows, rotation,
shocks, and turbulence, obtained by 
subtracting the thermal broadening from the observed H26$\alpha$ 
line width,  is less than 34 km s$^{-1}$.

For the  H66$\alpha$ line, the pressure broadening for 
gas with  ${\rm n_e}=1.1\times10^6$ cm$^{-3}$ accounts for 
23 km s$^{-1}$. Subtracting
$\Delta V_{\rm th}$, $\Delta V_{\rm nth}$ and $\Delta V_{\rm P}$
from the observed 
$\Delta V_{\rm FWHM}$ for the H66$\alpha$ line, 
 $\sqrt{\Delta V_{\rm FWHM}^2
-(\Delta V_{\rm th}^2 + \Delta V_{\rm nth}^2 + \Delta V_{\rm P}^2)}$ ~ $\approx
43$ km s$^{-1}$,  the residual line width, appears still to dominate 
the observed H66$\alpha$ line width, suggesting that the 
electron density estimated from the 7 mm continuum data \citep{dpre98}
might be underestimated due to the large optical depth of 
the free-free emission from the higher density gas. If 
the residual line width is all due to pressure broadening, 
the observed large H66$\alpha$ line width implies that a 
hyper-compact ionized core with an electron density  
$n_{\rm e}  {>} 2.5\times10^6$ cm$^{-3}$ is likely embedded in F3.

\begin{figure*}[t]
\centering
\includegraphics[width=145mm, angle=-90]{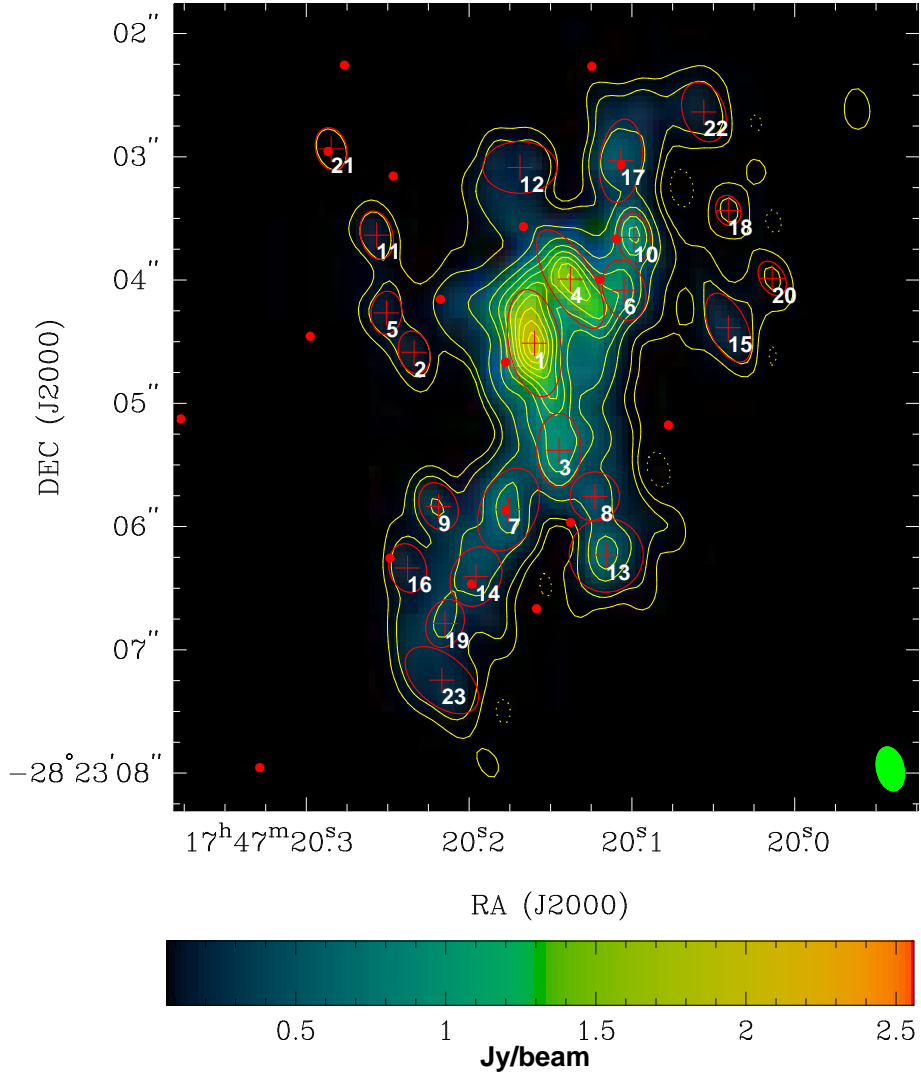}
\caption{
The image of 0.86-mm continuum emission observed with the SMA at a 
resolution of $0.36\arcsec\times0.22\arcsec$ (PA=13$\arcdeg$). The contours 
are (--0.02,0.02,0.05,0.15,0.25,0.35,0.45,0.55,0.65,0.75,0.85,0.95)$\times$2.34
Jy beam$^{-1}$. The color wedge shows the intensity range in units of
Jy beam$^{-1}$. The positions (cross) and the apparent shapes (ellipses)
of the continuum sources are derived from Gaussian fitting. 
}
\label{fig:dust}
\end{figure*}

\subsection{Continuum Emission}
The contours in Figure \ref{fig:H26aline} show
the continuum emission with the same FWHM beam 
($0.4\arcsec\times0.28\arcsec$) as the H26$\alpha$ 
line. The continuum flux densities corresponding to the 
H26$\alpha$ line emission regions are summarized
in column 8 of Table 2. A large fraction of the continuum emission 
at 0.86 mm appears to arise from a region outside the 
H26$\alpha$ line sources, revealing an overall extended 
filamentary structure northwest to southeast. Figure 
\ref{fig:dust} shows the details of the continuum 
emission at the higher resolution ($0.36\arcsec\times0.22\arcsec$). 
In addition to the UC HII regions observed at millimeter 
and centimeter wavelengths at the positions denoted with 
the red dots, we identified 23 
continuum sources at submillimeter wavelengths
which are listed in the bottom section of Table 2 and 
are named as Smm- with a sequential number (column 1) 
based on their angular distance $\Delta\theta$ from F3 
(column 7). The corresponding sequential numbers of the Smm- sources
are also marked in Figure \ref{fig:dust}. Using Gaussian 
fitting, we determined their positional offsets from F3 
in RA and DEC (columns 2 and 3), deconvolved sizes (column 4) and peak 
brightness and total flux densites (columns 5 and 6).
The brightness temperatures of the sources are given in column 8.
We note that the two brightest continuum cores are Smm-1 and Smm-4
with brightness temperatures 340 and 270 K, respectively.
Smm-1 is located close to but with a siganificant offset 
0.13\arcsec$\pm0.01$\arcsec\, from H26$\alpha$-1 which is associated
with F3. Smm-4 is located 0.06\arcsec$\pm$0.02\arcsec\, 
southeast of its  H26$\alpha$ counterpart  H26$\alpha$-4 
in the F1 complex.

%--------------------------------8 apr edits to here  ------ mchw ------------------
% edited revised version, v_r4to31may2011 MCHW 01 Jun  08:40 PDT

\section{Model for UC HII regions}
\begin{figure*}[t]
\centering
\includegraphics[width=130mm, angle=-90]{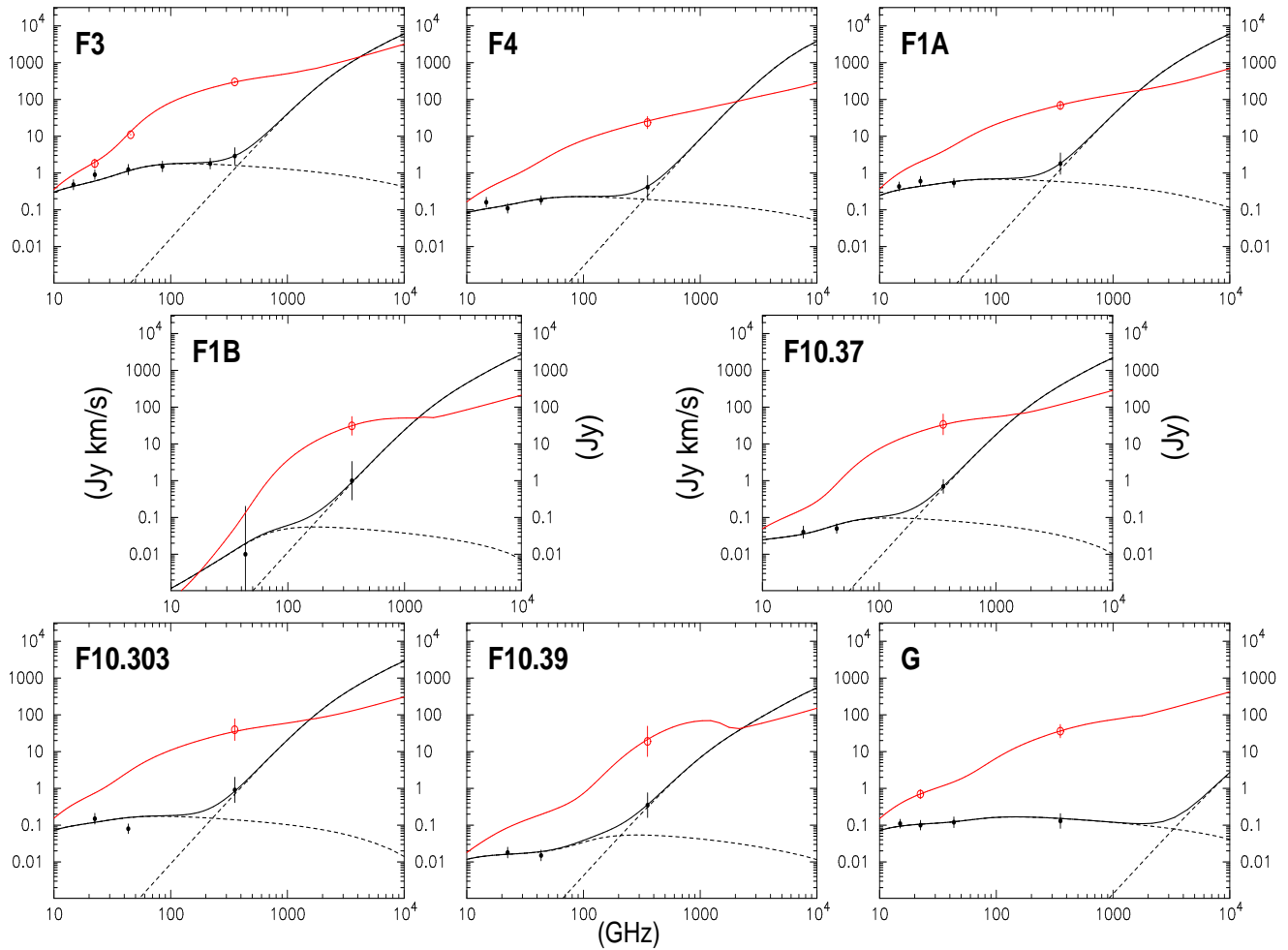}
\caption{Models for the UC HII regions F3, F4, F1A, F1B, F10.37
F10.303, F10.39 and G with the constraints from the observed
radio continuum flux densities and the fluxes of hydrogen 
recombination lines at radio and submillimeter wavelengths.
The red circles denote the integrated flux (1.064\,$S_p*\Delta V_{\rm FWHM}$) 
of the hydrogen recombination lines at H26$\alpha$ (from the 
SMA measurements of this paper), H52$\alpha$ and H66$\alpha$
from the VLA measurements \citep{dpre96}. The black dots are the 
flux densities determined
at 2, 1.3 cm and 7 mm with the VLA and 3 mm with the Hat Creek
interferometer array and 1.3 and 0.86 mm with the SMA (see Section 4.1,
the 1.3 mm data from Zhu 2010, personal communication).
The red curves indicate the fluxes of the hydrogen recombination lines
from the best fitted model (see section 4) to the observed data.
The black curves indicate the continuum spectral energy distribution
(SED) resultant from the free-free (flat dashed curves corresponding to
the first term in Equation (2) ) and dust
(steep rising curves corresponding to
the second term in Equation (2)) emissions 
calculated from the best fitted model.  
}
\label{fig:spfit}
\end{figure*}

On the assumption of isothermal homogeneous HII gas, 
a model has been developed for the low-frequency radio recombination
lines (RRL) including non-LTE effects, such as stimulated emission
by the background radiation \citep{shaver75}.  This model has been 
applied successfully  to the high-frequency radio recombination lines 
(H30$\alpha$) at 231.9 GHz from the high-density gas in the minispiral 
within the circum-nuclear disk of the Galactic center \citep{zhao10}. 
The effects of the radiation from both synchrontron and dust sources 
seem to be negligible for the H30$\alpha$ line under the conditions
within the central few parsecs of the Galaxy. In the massive
star forming cores such as Sgr B2 Main, a considerable fraction 
of the continuum flux density at submillimeter wavelengths is 
from dust radiation that needs to be included in the model. 
Using a homogeneous isothermal model with temperatures of $T_{\rm e}$ and 
$T_{\rm d}$ for both HII gas and dust, 
the flux densities of hydrogen recombination line ($S_{\rm L}$)  
and continuum ($S_{\rm C}$) emission at  radio to submillimeter 
wavelengths can be formulated assuming that the total 
dust is evenly distributed in three regions (each with an optical
depth $\tau_{\rm d}$) along the line-of-sight --- 
one third of the dust uniformly 
mixed with the ionized gas in a given HII region, and other two
thirds are located in front of and behind the HII region, respectively.
Equations (1) and (2)  give the full solutions 
for the line and continuum flux densities
from the three regions with no assumptions of small optical depths.
Spectral fits to the data are shown in Figure \ref{fig:spfit} .
The results are not significantly different from an optically-thin dust
approximation. This is because, in the centimeter to 
submillimeter wavelength range 
discussed in this paper, the dust emission towards the UC HII regions 
is indeed optically thin and  we are not able to discern between the 
two models. However, at mid-IR (Spitzer) wavelengths,  the dust 
becomes opaque and the dust attenuation  
towards the HII regions becomes important.

% I added the explanation in the referee response to the ms.tex mchw 21jul2011
  
\begin{eqnarray}
S_{\rm L}&=&{2k\nu^2\over c^2} \Omega\,T_{\rm e} 
\Bigg[\left(\tau_{\rm L}/\beta_{\rm N} +\tau_{\rm C} \over \tau_{\rm L} + 
\tau_{\rm C}^\prime \right)\times \nonumber \\ 
     & & \left(1-e^{\displaystyle -(\tau_{\rm L}+\tau_{\rm C}^\prime)}\right) 
-\left(1- e^{\displaystyle -\tau_{\rm C}^\prime}\right)
{\tau_{\rm C}\over\tau_{\rm C}^\prime}\Bigg]e^{\displaystyle -\tau_{\rm d}}+ \nonumber \\
   & &B(T_{\rm d})\tau_{\rm d}\Omega\Bigg[{{1-e^{\displaystyle -(\tau_{\rm L}+\tau_{\rm C}^\prime)}}
 \over{\tau_{\rm L}+\tau_{\rm C}^\prime}}-{{1-e^{\displaystyle -\tau_{\rm C}^\prime}}
             \over {\tau_{\rm C}^\prime}} \Bigg]e^{\displaystyle -\tau_{\rm d}}+ \nonumber\\
 & &B(T_{\rm d})(1-e^{\displaystyle -\tau_{\rm d}})\Omega
\Bigg[e^{\displaystyle -\tau_{\rm L}}-1\Bigg]e^{\displaystyle -(\tau_{\rm C}^\prime+\tau_{\rm d})}
  \, ,\\
\nonumber\\
{\rm and} & & \nonumber \\
\nonumber\\
S_{\rm C} &=& {2k\nu^2\over c^2}\Omega\,T_{\rm e}\left(1- e^{\displaystyle -\tau_{\rm C}^\prime}\right){\tau_{\rm C}
\over\tau_{\rm C}^\prime}e^{\displaystyle -\tau_{\rm d}} +\nonumber \\
    & &B(T_{\rm d})\tau_{\rm d}\Omega{ {1-e^{\displaystyle -\tau_{\rm C}^\prime}}\over 
{\displaystyle \tau_{\rm C}^\prime}}e^{\displaystyle -\tau_{\rm d}}+\nonumber \\
&&B(T_{\rm d})(1-e^{\displaystyle -\tau_{\rm d}})\Omega
\Bigg[1+e^{\displaystyle -(\tau_{\rm C}^\prime+\tau_{\rm d})}\Bigg]  \, ,
\end{eqnarray}

\noindent where $k$ is Boltzmann's constant, $c$ is the speed of light, 
and the total continuum 
optical depth in the HII region $\tau_{\rm C}^\prime=\tau_{\rm C}+ \tau_{\rm d}$
includes both the free-free  and dust contributions. 
The recombination line ($\tau_{\rm L}$) 
and free-free continuum ($\tau_{\rm C}$) optical depths of 
the radiation from an HII region 
are given by Equations (A3) and (A4) in Appendix A, respectively.
The values of the population departure coefficients ($b_{\rm N}$ and 
$\beta_{\rm N}$) were calculated using the non-LTE code of \cite{gord02} 
based on the analysis of hydrogen recombination lines at  wavelengths
from radio to submillimeters \citep{walm90}. 
The quantity $B(T_{\rm d})$ is the Planck function with a dust temperature
$T_{\rm d}$.   In Equation (1),
the first term associated with 
${\displaystyle 2k\nu^2\over \displaystyle c^2} \Omega\,T_{\rm e}$  
is the line emission from the HII gas attnuated by the foreground dust 
with optical depth $\tau_{\rm d}$;
the second term associated with $B(T_{\rm d})\tau_{\rm d}\Omega$
accounts 
for the absorption of the internal dust radiation by the HII gas  if
$\tau_{\rm L}\ge0$,  or the line emssion stimulated 
by the internal dust radiation if $\tau_{\rm L}<0$; and the third term
associated with $B(T_{\rm d})(1-e^{\displaystyle -\tau_{\rm d}})\Omega$
is the  absorption of background dust radiation by the HII gas  if
$\tau_{\rm L}\ge0$,  or the line emssion stimulated
by the background dust radiation if $\tau_{\rm L}<0$.
On the other hand,
for an isolated dust source,  
the dust continuum flux density is 
\begin{eqnarray}
S_{\rm d}    &=& B(T_{\rm d})(1-e^{\displaystyle -\tau_{\rm d}}) \Omega \,. 
\end{eqnarray}
The  optical depth $\tau_{\rm d}$ of dust continuum radiation
can be described as
\begin{eqnarray}
\tau_{\rm d} &=& \kappa_{0}({\nu\over\nu_0})^\beta M_{\rm d} D^{-2}\Omega^{-1}\, ,
\end{eqnarray}

\noindent where $\kappa_{\nu}=\kappa_{0}({\nu/\nu_0})^\beta$
is the dust opacity per unit dust mass, $ M_{\rm d}$ is dust
mass, $D$ is the distance, and $B_{\nu}(T_{\rm d})$ is the
Planck function with a dust temperature of $T_{\rm d}$.
In the calculations throughout this paper, we adopted 
$\kappa_0 = 1.06$ cm$^2$g$^{-1}$ at the reference frequency 
$\nu_0=230$ GHz calculated by \cite{osse94} for high-density 
gas $n_{\rm H}=10^7$ cm$^{-3}$ using the standard MRN model  
\citep{mrn77} with thin ice mantles. The physical parameters 
for a given HII region can be determined by fitting an isothermal, 
homogeneous model to the data at the submillimeter wavelengths 
of this paper and the data taken from previous observations at 
radio and millimeter wavelengths as discussed in the following 
section.

\subsection{Radio continuum spectrum \& RRL }
{\bf  F3} is the brightest UC HII region in the 
Sgr B2 Main core as observed at wavelengths of 2, 1.3 and 0.7 cm 
\citep{gaum90,gaum95,dpre98}.  {\bf  F3}, which is associated 
with the brightest H26$\alpha$ line source, H26$\alpha$-1,  
was detected at 2 cm with a peak flux density  0.138 Jy beam$^{-1}$ 
(with a circular beam of 0.3\arcsec\,), 
and a total flux density 0.48 Jy in an extended 
region of  size  0.7\arcsec, FWHM \citep{gaum90}. 
With a size  0.5\arcsec$\times$0.5\arcsec, FWHM, the 
peak intensity and total flux density 
at 1.3 cm are 0.22 Jy beam$^{-1}$ and 0.9 Jy \citep{gaum95}. 
Five  HII componenents have been detected at high-angular 
resolution (0.049\arcsec$\times$0.079\arcsec) with the VLA 
at 7 mm \citep{dpre98}. The authors show that the major 
component F3-d  has a shell- or 
ring-like morphology. Observations with the Hat-Creek 
interferometer array at 3.5 mm, \cite{liu99} showed that the peak 
emission of 1.51 Jy beam$^{-1}$ (FWHM = 1\arcsec$\times$0.5\arcsec) is 
coincident with F3 corresponding to a total flux density of 4.38 Jy for
F1, F2, F3 and F4. The high-resolution (0.64\arcsec$\times$0.31\arcsec) 
SMA image at 1.3 mm  shows that the peak emission of 1.8 Jy coincides 
with F3. The H66$\alpha$ line has been detected 
with 1.1\arcsec$\times$1.6\arcsec\ resolution 
at V$_{\rm LSR}=68\pm2$ \kms\, showing a 
line-to-continuum (L/C) ratio of 3$\pm$0.3 \% and $\Delta V_{\rm FWHM}=63\pm5$ 
\kms determined from integrated line profile \citep{dpre96}. 
At the lower resolution 2.9\arcsec$\times$1.6\arcsec,
the H52$\alpha$ line is towards the F cluster at
V$_{\rm LSR}=57\pm2$ \kms\, with L/C of 14$\pm$2 \% and 
$\Delta V_{\rm FWHM}=59\pm2$ \kms \citep{dpre96}. \cite{dpre96} 
noticed that the LTE electron temperature (T$^*_e$) derived from 
the high L/C value of the H52$\alpha$ line is significantly lower by a factor 
of 2 than that derived from the H66$\alpha$ line 
under the assumption of optically thin and LTE gas.
If the L/C ratios are uniform across the F complex,
the fluxes of the H66$\alpha$ and H52$\alpha$ lines from the dominant
source F3 can be estimated from the L/C values and the continuum 
flux densities at the corresponding wavelengths.

For the other seven regions associated with the
H26$\alpha$ line source, the continuum flux densities at 2, 1.3 and 
0.7 cm used in this paper are from \cite{gaum90}, \cite{gaum95} and 
\cite{dpre98}. The H66$\alpha$ line flux from the UC HII 
region G (H26$\alpha$-8) is estimated from the L/C ratio determined by 
\cite{dpre96} and the continuum flux density of \cite{gaum95}.
 
Figure \ref{fig:spfit} shows the  observed spectra of the 
continuum (black dots) and hydrogen recombination 
lines (red dots) at radio and submillimeter wavelengths.  

In order to fit to both the continuum and hydrogen recombination 
line data obtained from the high-resolution observations at
wavelengths in the range between 2 cm to 0.86 mm, we considered
a few possible models.  A model with an LTE approach and a single 
isothermal, homogeneous HII component  was rejected since 
the observed large H26$\alpha$ line flux requires large free-free 
continuum optical depths ($\tau_{\rm C}>1$) at longer millimeter 
and centimeter wavelengths for a single component model.
At these wavelengths, the dust absorption ($\tau_{\rm d}<<1$)
is negligible. Then, 
the predicted line fluxes at centimeter wavelengths for larger quantum-number 
transitions are much smaller than the observed values because 
of the exponential drop in the LTE line flux due to the large 
free-free optical depth $\displaystyle S_{\rm L}^*
=\frac{2k\nu^2}{c^2}\Omega 
T_{\rm e} (1-e^{\displaystyle -\tau_{\rm L}})e^{\displaystyle 
-\tau_{\rm C}}$. 

The observed line flux densities at centimeter wavelengths could 
be fitted with a single isothermal, homogeneous component by 
adjusting $n_{\rm e}$ and $T_{\rm e}$ if the stimulating effect 
in a non-LTE approach is considered. However, the flat continuum 
spectra at centimeter wavelengths seen in all the 
eight HII sources (Figure \ref{fig:spfit}) place a critical restriction 
on a model with a single isothermal, homogeneous component, 
requiring an additional lower-density component. Thus, we fit each 
of the eight observed hyper-compact H26$\alpha$ sources and the 
UC HII regions surrounding them with two components corresponding to
high- and low-density ionized gas with small and large sizes, A and B, 
respectively. Further assumptions used in the two-component model  
are the volume filling factors $f_{\rm V}=0.1$ and 0.5 for 
 components, A and  B, respectively. The effect of shadowing 
between the two components is negligible. 

% ---------------edits to here 14 apr  11:54 PST    mchw. on MacBook Pro ---------

\subsection{Hyper-compact HII component}
Figure \ref{fig:spfit} shows the results of model fitting the observed 
hydrogen recombination line  (red curves) and continuum flux densities
(black curves) from each of the eight UC HII regions with two isothermal, 
homogeneous HII components A and B.
The best fitted parameters are summarized 
in Table 3. The first row gives the size in units of 10$^3$ AU.
The following rows under the category of HII gas properties summarize
the derived parameters including the temperature $T_{\rm e}$, 
density $n_{\rm e}$, volume filling factor $f_{\rm V}$, 
the departure coefficients $b_{\rm N}$ and $\beta_{\rm N}$, 
the optical depths of the H26$\alpha$ line $\tau_{\rm L}({\rm H26\alpha})$,
the free-free continuum  $\tau_{\rm C}({\rm 0.86\,mm})$ at 0.86 mm,
the fractional contribution $\zeta_{\rm H26\alpha}$ to the observed 
H26$\alpha$ line flux from each component,
the H66$\alpha$  line $\tau_{\rm L}({\rm H66\alpha})$, and
the free-free continuum  $\tau_{\rm C}({\rm \lambda13\,mm})$ at 13 mm.
Components A have small size (160 to 2000 AU), high electron 
density (3 to 33 $\times10^6$ cm$^{-3}$), relatively lower
temperature (5 to 9$\times10^3$ K). The A components are typically 10 
times smaller but 100 times denser than  typical UC HII regions,
$10^{17}$ cm and 10$^4$ cm$^{-3}$, \citep{chur02}, representing
a class of hyper-compact HII components \citep{chur02} present in the 
Sgr B2 Main core.
The hyper-compact HII components could arise from plausible ionized 
disks with emission measure (EM) in the range 0.4 to 
8.4$\times10^{10}$ cm$^{-6}$ pc, accounting for the most of 
the H26$\alpha$ line flux, {\it e.g.} $\sim$94\% for F3.
The hyper-compact HII components in all the HII cores show
a negative optical depth for the H26$\alpha$ line, suggesting the
presence of stimulated emission by free-free continuum emission
from the hyper-compact HII region itself. For F3,
the H26$\alpha$ line is enhanced by a factor of 1.4 due to 
a weak stimulating process
within the hyper-compact HII component (A) while the 
enhancement factors ($\displaystyle S_{\rm L}/S_{\rm L}^*$) 
are 1.7  and 4 for the H26$\alpha$ lines from 
G and F10.39 which show larger negative line optical depths of
$-0.71$ and $-1.3$, respectively. At 13 mm, the continuum component A
becomes optically thick while the H66$\alpha$ line
is still optically thin, $\tau_{\rm C}({\rm 13mm}) =9.5$ 
and $\tau_{\rm L}({\rm H66\alpha})=0.13$ for F3. Component A
appears to make a little contribution to the total line flux
of the H66$\alpha$ transition ($\sim$20\% for F3).

\subsection{Ultra-compact HII component}
The continuum emission at longer wavelengths from
each of the UC HII regions shows a shallow rising spectra 
towards short wavelengths, indicating the presence
of a larger HII component which might result from
an ionized stellar wind or expansion of an ionized nebula. 
Component (B) with a relatively large size (1200 to 4000 AU),
lower density (0.1 to 3.5$\times10^5$ cm$^{-3}$), and higher 
temperature (10 to 17$\times10^3$ K) is needed for the
extended ionized gas with relatively smaller emission measures
of $\le 1\times10^9$ cm$^{-6}$ pc.
 $T_{\rm e}=17,000$ K for the UC HII component in F3 (H26$\alpha$-1)
appears too high for the HII regions with  metallicities in the
Galactic center. The over-estimated $T_{\rm e}$ from our fitting might 
be caused by the under-estimate of the H66$\alpha$ and H52$\alpha$ 
line fluxes at 1.3 and 0.7 cm due to large line broadening and limited 
bandwidth coverage of the old VLA system. 
For a canonical value of $T_{\rm e}=10,000$ K, 
the corresponding  H66$\alpha$ and H52$\alpha$ line fluxes require
50\% more than the values used in this paper.
The physical parameters 
of the B components fall into the category of UC HII region in
\cite{chur02}. The continuum emission from the B components 
is optically thin at longer millimeter wavelengths, providing 
a flat spectrum ($\propto\nu^{-0.1}$) in contrast to the steep 
rising spectrum ($\propto\nu^{2}$) of component A. With its 
larger size, the free-free emission from the lower density gas 
(component B) dominates the continuum flux densities at 13 mm. 
Superposition of the spectra from the two components (A and B) 
results in a spectrum slowly rising at long wavelengths and 
turning over to flat at short millimeter and sub-millimter 
wavelengths.

% -----------edits to here 14 apr  14:54 PST    mchw. on MacBook Pro ---------
% edited revised version, v_r4to31may2011 MCHW 01 Jun  09:40 PDT

\subsection{Dust component} 
The composite spectra (dashed flat curves in Figure \ref{fig:spfit})  
of  components A and B appear to fit the observed continuum flux 
densities at millimeter and centimeter wavelengths. In the submillimeter, 
the observed flux densities show a significant excess emission from 
the UC HII region over the ones predicted from the isothermal, 
homogeneous ionized gas model with two-density components, 
suggesting that dust radiation becomes significant in the 
continuum flux density. In order to evaluate the contribution 
of  dust radiation in the UC HII regions, the terms involving 
the dust radiation $B(T_{\rm d})$ in Equations (1) and (2) are added to
the B component model with the dust distribution as assumed.   
The flux density 
($S_{\rm d}$) of the dust radiation is determined using Equations 
(3) and (4) with an assumed dust temperature in the HII 
region equal to the peak temperature in the Sgr B2 Main core, 
$T_{\rm d}\sim500$ K, \citep{lis90} and a power-law index for the dust
opacity $\beta=1.5$ \citep{gold90,lis90}. The observed brightness
temperatures of few times $10^2$ K for the continuum emission from
 several central dense clumps also suggest a high dust temperature 
for the UC HII regions in Sgr B2 Main (also see the discussion 
in Section 5 for the high dust temperature). 

We show that, except for G, the dust radiation in  the other seven regions 
makes a contribution to the observed continuum flux density at 0.86 mm
comparable or larger than that from free-free emission
(see Figure \ref{fig:spfit}).  In general, the dust in   
each of the UC HII regions is optically thin, $\tau_{\rm d}\sim ~ 0.1$, 
with a total gas mass  $\le 11 M_\odot$.          
The dust optical depth and the gas mass ($M_{\rm tot}$)
needed in each of the UC HII regions are summarized  in Table 3. 

Finally, the models with two independent HII components mixed with dust 
involved a set of ten free parameters, namely 
2$\times$($n_{\rm e}$, $T_{\rm e}$,
R, $f_{\rm V}$, and $T_{\rm d}$). For F3, we have collected a total
of nine measurements in line and continum flux densities from 
high-resolution observations at centimeter-submillimeter wavelengths.
Adding the two determined angular sizes for the components A and B,
we have a total of eleven measurements to constrain
the model reasonably well. However, for the rest of the sources, the models
need to be confined with further high-resolution observations
of both continuum and line emissions at wavelengths from  
centimeters to submillimeters.  

% edited revised version, v_r4to31may2011 MCHW 01 Jun  10:40 PDT

\subsection{Lyman continuum}
The hyper-compact HII components detected with H26$\alpha$ lines
appear to require most of the flux of Lyman continuum photons 
from the ionizing stars to maintain their ionization.
The H26$\alpha$ line flux from hyper-compact sources would
be excellent tracers for newly formed massive stars.
Given an observed FWHM linewidth of $\Delta V_{\rm FWHM}$ that
is dominated by Doppler broadening $\Delta V_{\rm D}$ in the 
H26$\alpha$ line profile, the ionizing photon flux required for the
observed  H26$\alpha$ line flux ($S_{\rm L}$) can be estimated from 
Equation (A8) that was generally derived for a radio recombination line 
${\rm N+1\rightarrow N}$,

\begin{eqnarray}
N_{\rm Lym}
 &=& 7.8\times10^{46}\left({S_{\rm L}\Delta V_{\rm D}}\over{\rm Jy~km~s^{-1}}\right)
\left({\nu_{\rm H26\alpha}\over\nu}\right)\left({D}\over{\rm 8\,kpc}\right)^2\times \, \nonumber \\
            &&\left({T_{\rm e}}\over{10^4\,K}\right)^{1.5}
       \left(\alpha_{\rm B}(T_{\rm e}) \over\alpha_{\rm B}(10^4\,K)\right)
     \Psi^{-1}_{26} ~~~~~~~~~ph~s^{-1} \,
\end{eqnarray}
where $\alpha_{\rm B}(T_{\rm e})$ is the total recombination coefficient to 
excited levels \citep{humm63}, and  the function
$\Psi_{26}$ provides a correction 
for the effects of both non-LTE and optical-depth. 
The values of $\Psi_{26}$ are
given in Table 3 for each case. For the hyper-compact 
HII components (A) in Sgr B2 Main, $\Psi_{26}$ is in the range  
$\sim$~1 (H26$\alpha$-6) to $\sim$~2 (H26$\alpha$-7).
For the UC HII components (B) where the H26$\alpha$ line 
is optically thin, $\Psi_{\rm 26}\approx\rm b_{26}$, 
a correction for the factor due to the
lower quantum-number levels that are underpopulated with respect to the LTE
in the cases of lower electron density. The electrons at the level of ${\rm N=26}$
appear to be underpopulated by 20\% to 30\% with respect to the LTE
for the UC HII regions in Sgr B2 Main.

In addition, 
for  H26$\alpha$-1 (F3) assuming
$T_{\rm e}=9,000$ K, the value $N_{\rm Lym}=1.43\times10^{49}$
ph s$^{-1}$ is inferred from the
observed H26$\alpha$ line flux 280 ${\rm Jy\,km\, s^{-1}}$
using Equation (5)  with $\Psi_{26}=1.3$ while 
$N_{\rm Lym}=2.66\times10^{49}$ ph s$^{-1}$ is
derived from the continuum flux density $S_{\rm C}=2.9$ Jy at 353.6 GHz using
Equation (4) given 
by \cite{wilc94}.
The value determined from the observed continuum flux density
surpasses the
value determined from H26$\alpha$ line flux by 46\%.
Our detailed model fitting gives the free-free flux density
of 1.61 Jy suggesting the fraction of dust contribution to the
continuum is 44\% at 353.6 GHz, in good agreement with above
assessment.
Considering optically thin of the line emission and
nearly no attenuation and contamination by the dust
at 353.6 GHz (0.85 mm), it appears to be an excellent way
to determine the $N_{\rm Lym}$ and the free-free flux density
using the H26$\alpha$ line flux.
Therefore, comparing the observed continuum flux density
with the free-free flux density determined from the H26$\alpha$ line flux,
one can give a good assessment of the fraction of
dust contribution to the continuum.

% -----------edits to here 14 apr  16:00 PST    mchw. on MacBook Pro ---------

\subsection{Ionizing stars}
Based on the derived properties for both A and B components, we estimated 
the fluxes of the Lyman continuum photons ($N_{\rm Lym}$) from the newly 
formed ionizing stars. In each of the UC HII regions, the hyper-compact 
% changed ultra -> hyper above. mchw.
component (A) appears to require more flux in Lyman continuum photons 
than their larger but lower-density counterpart (component B) for the 
maintenance of the ionization (see Table 3). A total flux in Lyman 
continuum photons ($N_{\rm Lym}^{\rm A+B}$) is evaluated by the addition 
of the individual fluxes required for components A and B. Assuming that 
$N_{\rm Lym}^{\rm A+B}$ accounts for all the Lyman continuum photons 
from a single early-type, zero-age-main-sequence (ZAMS) star in each 
of the UC HII regions with no significant leakages, {\it i.e.} the 
UC HII regions are internally ionized \citep{dpre98}, the type of  
massive stars required for each of eight HII complexes with a 
hyper-compact H26$\alpha$ component is inferred on the basis
of the stellar atmosphere model computed by \cite{pan73} and listed in 
Table 3. We note that the Lyman continuum photon fluxes from \cite{pan73}'s 
stellar atmosphere model were underestimated by 26\% to 66\% for O6 to B0 type stars,
respectively, as compared to those computed from \cite{Vacc96}'s model.
In good agreement with \cite{dpre98}, we also find that 
at least an O6 star is required for the brightest H26$\alpha$ source 
(H26$\alpha$-1) in F3 and an O8.5 star for H26$\alpha$-2 in F4. 

The complex F1, including H26$\alpha$-3 (O7), H26$\alpha$-4 (O9.5) and  
H26$\alpha$-6 (O9), has been resolved by the VLA with resolution 
of 0.049\arcsec$\times$0.079\arcsec\, into at least seven compact components 
\citep{dpre98}. The authors suggest that the complex requires a group
of seven early-B and late-O type stars (B0 to O8.5) to maintain the ionization.
The flux of Lyman continuum photons ($N_{\rm Lym}$) inferred 
from our analysis of the H26$\alpha$ line for the hyper-compact sources in 
F3 and F1 appear to be considerably greater than the values inferred from
their 7 mm-counterparts based on the VLA continuum observations \citep{dpre98}.
The difference occurs because the hyper-compact components
H26$\alpha$-1  and H26$\alpha$-3 in F3 and F1 are optically thick in 
free-free emission at 7 mm, $\tau_{\rm C}\approx2$, inferred from
the analysis above on the basis of H26$\alpha$ observations. 
The hyper-compact core appears to be deeply embedded in the optically 
thick region at 7 mm. From the VLA flux densities at 7 mm, the flux of 
Lyman continuum photons has possibly been underestimated due to 
missing the contribution from the hyper-compact ionized core. 
Thus, an O7 star might be needed to maintain 
the ionization of hyper-compact H26$\alpha$-3 in the HII complex of F1.   
  
The above argument is also valid for the 
hyper-compact HII components H26$\alpha$-5 (O9), H26$\alpha$-7 
(O9.5) and H26$\alpha$-8 (O8.5). These hyper-compact components appear 
to be embedded in the relatively isolated UC HII regions F10.37, 
F10.39 and G, respectively. In fact, the continuum optical depths inferred for 
the three hyper-compact HII components are greater than the values 
derived for the rest of hyper-compact HII components (see Table 3).

As discussed above (see Figure \ref{fig:spfit}), the H26$\alpha$ line from
the three components appears to be enhanced by stimulated emission from 
continuum emission in the hyper-compact HII gas surrounding the newly 
formed O stars. The actual flux of the ionizing photons is reduced 
by a factor of 4 in H26$\alpha$-7 as compared to an LTE source. 
Thus, the effect of stimulated line emission helps detection of H26$\alpha$ 
line from a region ionized by an O9.5 
among the eight H26$\alpha$ line sources in SgrB 2 Main.

Given a hyper-compact HII region with $T_{\rm e}=1\times10^4~K$, 
$\Delta V_{\rm D}=30$ km s$^{-1}$ without line stimulation,
the 3$\sigma$ (0.3 Jy beam$^{-1}$) detection limit for
the H26$\alpha$ line imposed by the SMA data used in this paper gives 
a limit of $N_{\rm Lym}=7.1\times10^{47}$ ph s$^{-1}$, or
$\log\left(N_{\rm Lym}\right)=47.85$, on
the flux of ionizing photons, corresponding to an O9.5 ZAMS star.
No significant detections of the H26$\alpha$ line with the SMA towards F2 are
consistent with  
four B0 stars inferred from the 7-mm continuum observations \citep{dpre98}.

\subsection{Ages}
The ages of the newly formed O stars in Sgr B2 Main can be
assessed by the dynamic time-scale of the larger HII components (B)
as the time of sound wave traveling from the initial
ionization front at a radius close to the Str\"omgren radius
($R_{\rm S}$) \citep{Str39} as suggested by \cite{spiz78} and \cite{GL99}. 
Using the Equation (3) given in \cite{SHI10b} and the isothermal 
sound speed $C_{\rm S}=\sqrt{kT_e/m_{\rm H}}$, we calculated the 
dynamical age of the HII components (B) and listed the values 
in Table 3, ranging between the oldest one, 1.3$\times10^3$ y
for F3 (H26$\alpha$-1) and the youngest one, 0.4$\times10^3$ y 
for  F10.37 (H26$\alpha$-7). 
If there is any external pressure produced by the swept-up ISM as the
nebula expands, the sound crossing time will give an under-estimate of
the age. The O type stars in Sgr B2 Main 
appear to be clustered  around a thousand years ago. 
 
\section{A cluster of protostellar cores}
The dust emission from Sgr B2 Main has been resolved into at least 
twenty three components, showing that a cluster of protostellar cores
are present in this region in addition to the newly formed  O and B 
type stars suggested by SMA observations of the H26$\alpha$ line 
and VLA observations of the radio continuum emission and
H52$\alpha$ and H66$\alpha$ lines. From our interferometer 
observations, the brightness temperature ($T_{\rm b}$) of the 
continuum emission from individual components can be determined
under the Rayley-Jeans approximation,
\begin{equation}
T_{\rm b}(\nu) \approx 
13.6\lambda_{\rm mm}^2\left[\frac{S_{\rm d}}{Jy}\right]
\theta_{\rm FWHM}^{-2}~~~K, 
\end{equation}
where $S_{\rm d}$ is the dust flux density at wavelength, $\lambda$ 
from the region with the geometric mean of the angular size 
$\theta_{\rm FWHM}$ in arcsec. Located near (0.13\arcsec$\pm0.01$\arcsec 
north) the brightest H26$\alpha$ line source in F3, Smm-1 has 
the highest brightness temperature of 340 K. The  contribution 
to the brightness temperatures in the continuum cores from the free-free 
continuum emission is less than 10\%.  Thus, the brightness temperature 
from dust emission is greater than 300 K in the center, 
suggesting that high dust temperature is present in Sgr B2 Main.
 
 % -----------edits to here 14 apr  16:30 PST    mchw. on MacBook Pro ---------

From Equations(3) and (5), for a given dust temperature $T_{\rm d}$,
we can determine the optical depth  ($\tau_{\rm d}$) for each of the dust
cores from $T_{\rm b}(\nu)$:
\begin{equation}
    \tau_{\rm d}(\nu) = {\rm log}\left[\frac{T_{\rm d}}{T_{\rm d}-T_{\rm b}(\nu)}\right]
\end{equation}
The dust temperature ($T_{\rm d}$) is affected by the increasing luminosity
in the inner region ($\sim$0.1pc) of the Sgr B2 Main core where at least eight newly
formed O stars are  suggested by the H26$\alpha$ line 
sources. We determined the physical properties of the proto-stellar 
cores using a power-law distribution for  the mean dust temperature 
$T_{\rm d}$ as function of the radial distance $r$ from the center. 
The distribution of $T_{\rm d}$ is modeled by \cite{scov76} and  \cite{wolf87} assuming 
that the total emissivity of the collection of grains at that temperature 
equals the total radiative energy that is absorbed by the grains 
\citep{scov76, wolf86}, 
\begin{equation} 
T_{\rm d} = T_{\rm in}\left(\frac{r}{r_{\rm in}}\right)^{-\Gamma}
\end{equation}
where  $T_{\rm in}$ is the dust temperature at $r_{\rm in}=0.01$ pc,
the radial distance between Smm-1 and H26$\alpha$-1 (O6 star) assuming a
mean projection angle of 45\arcdeg.   Depending on the dust emissivity 
($\rm Q_\nu$), the central stellar luminosity is insensitive to 
the exponent $\beta$ of the dust emissivity power-law ($\Gamma=2/5$ or $1/3$ 
for $\beta=1$ or 2, respectively). 

We used a linearly interpolated value of 
$\Gamma=0.37$ for  $\beta=1.5$ in the modeling.  Assuming  
dust heated primarily by the central O6 star with a luminosity of 
2.5$\times10^5\, L_\odot$, $T_{\rm in}(r_{\rm in}=0.007 
\,{\rm pc})\approx$400 K is inferred from the structure of the dust 
temperature derived by \cite{scov76}. On the other hand, a larger 
value of $T_{\rm in}(r_{\rm in}=0.007 \,{\rm pc})\approx630$ K is 
calculated by \cite{lis90} in their detailed models assuming a 
total luminosity up to 2.3$\times10^7\, L_\odot$ from the stars
distributed in the core in Sgr B2 Main and exponent of $\beta=1.5$ 
in the dust emissivity power-law. Because of high  infrared opacity 
in the inner region of the core, the radiation from the
outer region does not affect the dust temperature at the center.
The observed brightness temperature, $T_{\rm b}=300$ K from the dust
in Smm-1 indicates the dust temperature $T_{\rm d}\approx470$ K if 
 $\tau_{\rm d}\approx1$, which is consistent with the kinetic 
temperature of the absorbing gas towards Sgr B2 Main 
\citep{cern06,qin08}. We note that the dust 
sublimation radius  $R_{\rm sub}$ by  heating from the central 
star can be  determined by balancing the dust absorption of the UV 
and visible radiation from the star with the re-emission at IR from the 
dust grains at its sublimation temperature  
$T_{\rm sub}=1800$ K \citep{wolf86}. For an O6 star, 
$R_{\rm sub}=2.9\times10^{15}$ cm, is about an order magnitude 
smaller than $r_{\rm in}$ and in the zone between the two 
radial distances, the dust substantially cools down due to 
infrared radiation. In the following calculation, 
we take $T_{\rm in}=500$ K in the power-law exponent of 
$\Gamma=0.37$ for the dust temperature. 
Using Equation (8), we calculate  $T_{\rm d}$ in the range 
500 K for Smm-1 near the O6 star to 170 K for Smm-23 about 0.14
pc away. Using Equations (6), (7) and (8), we find that dust 
optical depth lies between 0.3 and 3.0 at 0.86 mm for these  
protostellar cores, with unity for the brightest core Smm-1.
Table 4 summarizes the optical depths ($\tau_{\rm d}$) 

From Equations (9), (10) and (11), we can calculate the 
total gas masses ($M$), surface density ($\Sigma$) and 
molecular number density ($n_{\rm H}$) in the individual protostellar
cores assuming a mass ratio of gas-to-dust $g=100$ and
a mean mass per molecule, $\mu=2.3$,
\begin{eqnarray}
M  &=&  12.0  
\left[\frac{\nu}{\rm 230 GHz}\right]^{-\beta}
{\rm log}\left[\frac{T_{\rm d}}{T_{\rm d}-T_{\rm b}(\nu)}\right] 
\times
\nonumber \\
 &&
 D_{\rm kpc}^2  \theta_{\rm FWHM}^2~~~{M_\odot}  \, , 
\end{eqnarray}
\begin{eqnarray}
\Sigma &=&94.3 \left[\frac{\nu}{\rm 230 GHz}\right]^{-\beta}
{\rm log}\left[\frac{T_{\rm d}}{T_{\rm d}-T_{\rm b}(\nu)}\right]
g~{\rm cm^{-2}} \, ,
\end{eqnarray} 
and 
\begin{eqnarray}
n_{\rm H}&=& 16.4\times10^8 \left[\frac{\nu}{\rm 230 GHz}\right]^{-\beta} 
 {\rm log}\left[\frac{T_{\rm d}}{T_{\rm d}-T_{\rm b}(\nu)}\right]
\times
\nonumber \\
 &&D_{\rm kpc}^{-1}  \theta_{\rm FWHM}^{-1}~{\rm ~cm^{-3}} \, .
\end{eqnarray}

 % -----------edits to here 14 apr  17:00 PST    mchw. on MacBook Pro ---------

Table 4 lists the derived values for $M$, $\Sigma$ and 
$n_{\rm H}$ in columns (6), (7) and (8). 
The model predicts a mass 143 M$_\odot$ 
for  Smm-1, while the largest mass of 223 M$_\odot$ is 
associated with  Smm-4 near F1 (H26$\alpha$-3). The enclosed gas 
masses ($M$) of the individual dense cores range 5 to  223 
$M_\odot$. The molecular surface density is in the range 
13 to 150 $g\,{\rm cm^{-2}}$. The molecular number density is in the 
range  1$\times10^8$--1$\times10^9$ cm$^{-3}$, which appears to be 
consistent with the ambient density of $10^7$--$10^8$ cm$^{-3}$ 
suggested by \cite{dpre98} if ionized gas in the UC HII regions and 
the ambient molecular gas are in pressure equilibrium \citep{xie96}.

In order to evaluate whether the individual dust cores are subject to 
inevitably undergoing gravitational collapse to form stars, we calculated 
the Bonnor-Ebert mass ($M_{\rm BE}$), the maximum mass that a dense core 
remains in hydrostatic equilibrium \citep{bonn56,eber57}, and free fall 
time scale ($t_{ff}$), 
\begin{eqnarray}
M_{\rm BE} = 1.18 C_{\rm S}^3G^{-3/2} \rho^{-1/2} \,
\end{eqnarray}

\begin{eqnarray}
t_{ff} = \frac{1}{4}\sqrt{\frac{3\pi}{2G\rho}}
\end{eqnarray}
where $C_{\rm S}$, $G$ and $\rho$ are the sound speed, gravitational 
constant and mass density of the core. The derived values for 
$M_{\rm BE}$ and $t_{ff}$ are listed in columns 9 and 10 of Table 4. The masses 
of the protostellar cores in the central region of Sgr B2 Main are 
1 to 2 orders of magnitude greater than their corresponding Bonnor-Ebert 
mass. Free-fall time scales of 1--3 $\times10^3$ y are inferred 
from our model calculation.

\begin{figure*}[t]
\centering
\includegraphics[width=125mm, angle=-90]{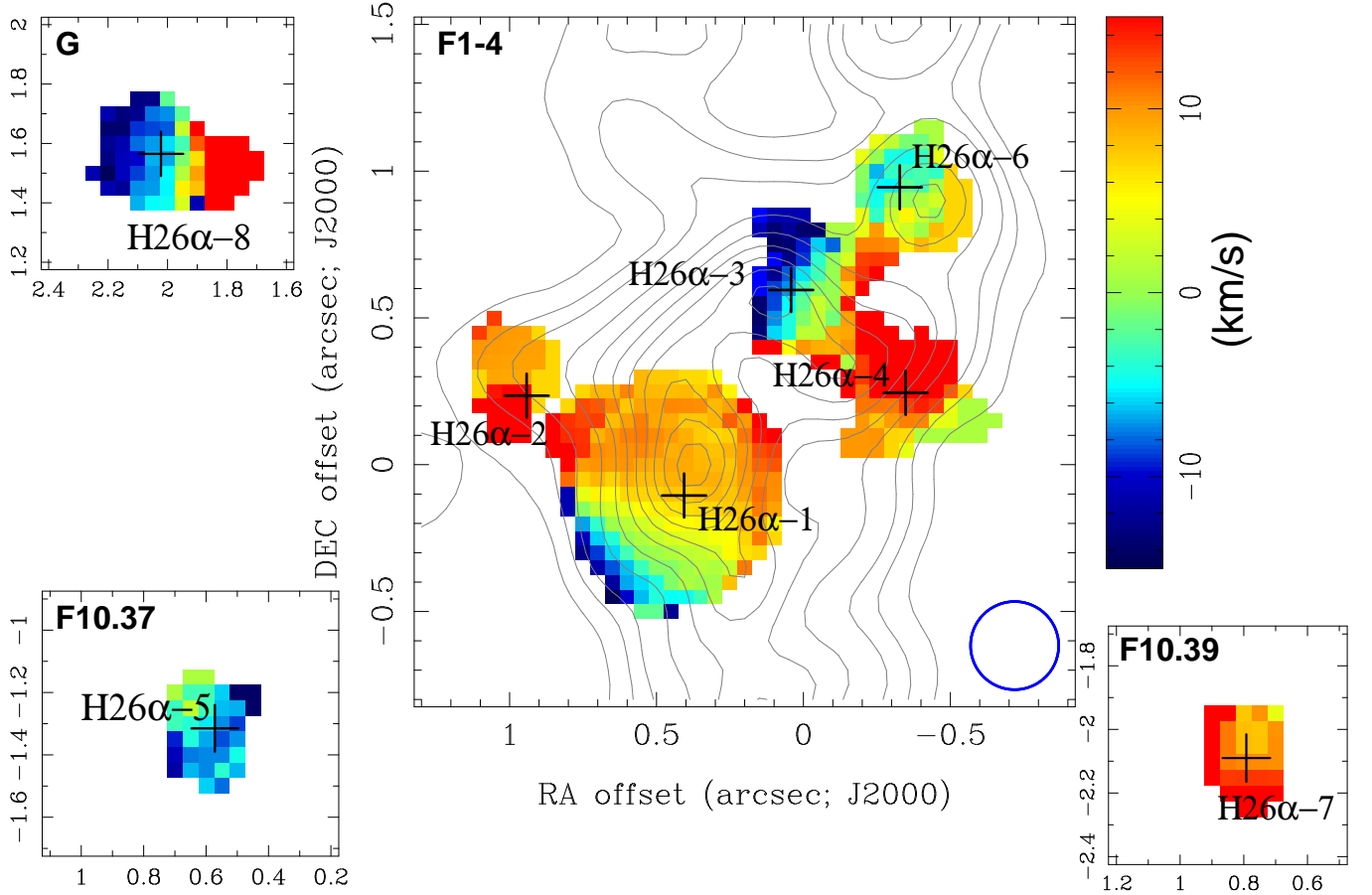}
\caption{
The distribution of intensity weighted radial velocity with respect to a 
LSR velocity of 58 km s$^{-1}$ is overlaid on the contour image of the 
continuum emission at 0.86 mm (see Figure \ref{fig:H26aline}). The velocity 
image was constructed from the H26$\alpha$ line image cube convolved with 
a circular beam of 0.3\arcsec\, using the moment 1 algorithm in Miriad with 
4$\sigma$ intensity cutoff. The contours are (0.25,0.35,0.45,$\dots$,0.95)
$\times$ 2.5 Jy beam$^{-1}$ with a circular beam of 0.3\arcsec. The cross 
signs mark the peak positions of the H26$\alpha$ line souces of 1, 2, 3, 4 
and 6 in the central F1-4 complex (the main figure) along with the 
sources of H26$\alpha$-5, 7, and 8 in the discrete sources with 
large angular offsets from the center (the insets). Both the vertical 
and horizontal coordinates are the offsets (arcsec) from the registration 
center: R.A.(J2000) =17:47:20.135 and Decl.(J2000) =--28:23:04.53. 
}
\label{fig:kin}
\end{figure*}
\section{Discussion}
The kinematics of the F1-4 HII complex and the discrete HII regions
in Sgr B2 Main are shown by the radial velocity distribution
determined from the first moment images of the H26$\alpha$ line in
Figure \ref{fig:kin}. The observed kinematics imply certain dynamical 
processes in the ionized gas associated with the newly born massive stars 
in the complex. Most of the H26$\alpha$ line emission arises from the 
high-density component (A) in an H26$\alpha$ source in Sgr B2 Main, 
as shown with our model fitting (see Table 3). The high-density, 
hyper-compact components with a typical size of a few $\times10^2$ AU 
are  candidates for rotating ionized disks.

\subsection{Candidate for rotating disk}
\noindent {\bf H26$\alpha$-7 in F10.39}, a weak stimulated source, shows a 
radial gradient, nearly north-to-south, with radial velocity difference
of 9 km s$^{-1}$ or $ V_{\rm r}\approx4.5$ km s$^{-1}$ across the central 
0.2\arcsec\ , or $2R=1600$ AU, which could indicate 
 an edge-on rotating disk (see Figure \ref{fig:kin}). If the velocity 
difference is due to Keplerian rotation of the ionized gas in a disk 
with an inclination angle $i$, then the enclosed mass is
\begin{eqnarray}     
M &\approx& 1.1\,M_\odot\left(V_{\rm r}\over {\rm km~s^{-1}}\right)^2 \left(R\over {\rm 10^3AU}\right){\rm sin}^{-2}(i)
\, \nonumber \\
   & \approx& 18\,{\rm sin}^{-2}(i)~~~~M_\odot,  
\end{eqnarray}
which is consistent with the mass of a late O-type star as 
the ionizing source. However, we note that the observed kinematics
of the two sources can also be interpreted as a rotating-expanding HII shell.
The data are not adequate to discriminate between disk and rotating-expanding
shell models.
  
\subsection{Confined HII ouflow and expanding ionized ring or disk}

\noindent {\bf F3 (H26$\alpha$-1)}  shows a large velocity gradient 
 northwest-to-southeast (see Figure \ref{fig:kin}). 
The maximum radial velocity difference
is 50 km s$^{-1}$ or  $V_{\rm r}\approx25$ km s$^{-1}$ across a 
projected angular size of 0.7\arcsec\, or $2R= 5600$ AU which 
is consistent with the angular size of the ring/disk seen in the 
VLA high-resolution (0.049\arcsec$\times$0.79\arcsec) observation
at 7 mm \citep{dpre98}. The ring/disk is nearly circular with a
ratio of minor-to-major axis size $\theta_{\rm min}/\theta_{\rm maj}\gtrsim
0.5\arcsec/0.7$\arcsec, indicating an inclination angle $i\lesssim$ 45\arcdeg.
If the velocity difference is due to Keplerian rotation of the ionized gas
in the ring, then from Equation (14), an enclosed, dynamic mass 
$1.9\times10^3\, {\rm sin}^{-2} (i)~M_\odot$ can be inferred,
which is at least an order magnitude greater than the inferred
total mass of both the ZAMS type O star ($\sim$30$\,M_\odot$) in H26$\alpha$-1
and the protostellar core Smm-1 
($\sim$34$\,M_\odot$) in F3 (see Tables 3 and 4).
On the other hand, the electron temperature of $T_e=1.7\times10^4\, K$, 
inferred for the large, lower-density component
(B) from our model fitting, suggests an isothermal 
sound speed  $C_{\rm S} = \sqrt{kT_{\rm e}/m_{\rm H}} =12$ km s$^{-1}$ which is
about a factor of three smaller than the maximum expansion velocity
$V_{\rm max} =  V_{\rm r} \,{\rm sin^{-1}(i)}\approx 35$  km~s$^{-1}$.
However, the radiation pressure 
($\displaystyle P_{\rm rad}=\frac{L_*}{{\rm 3c} \pi R^2 }$) 
due to a luminosity
$L_*$ from an O type star 
would be  comparable to the thermal pressure
($P_{\rm th}=2n_{\rm e}kT_{\rm e}$) in the HII region with 
a radius of $R$, where
the factor of 2 accounts for equal densities of 
electrons ($n_{\rm e}$) and protons ($n_{\rm i}$), 
\begin{eqnarray}
\frac{P_{\rm rad}}{P_{\rm th}} &=& 2.3\left(L_*\over 10^5L_\odot\right)
\left(R\over {\rm 10^3 AU}\right)^{-2}\times\, \nonumber \\
      &&\left(n_{\rm e}\over 10^6\,{\rm cm^{-3}}\right)^{-1}
\left(T_{\rm e}\over 10^4 K\right)^{-1},
\end{eqnarray}
where $\displaystyle \frac{P_{\rm rad}}{P_{\rm th}}$
is the  ratio of the radiation pressure 
to the thermal pressure.
On the basis of the parameters derived for F3
(see Table 3), $\displaystyle \frac{P_{\rm rad}}{P_{\rm th}}\approx0.4$  
and 0.7 for the high-density component (A) and the lower-density component (B), 
respectively. On the other hand, if the radial velocity gradient is 
mainly due to expansion or outflow, the ram pressure of the expansion 
motions is $\rho_{\rm HII}V_{\rm ex}^2$. Then, the ratio of the ram 
pressure to the thermal pressure is,
\begin{eqnarray}
\frac{P_{\rm ram}}{P_{\rm th}} &=& 6.1\times10^{-3}\left(T_{\rm e}\over10^4\,K\right)^{-1} 
\left(V_{\rm ex}\over {\rm km\,s^{-1}}\right)^2 .
\end{eqnarray} 
For the lower-density component (B) of F3, 
$\displaystyle \frac{P_{\rm ram}}{P_{\rm th}} \approx0.5$ 
if $V_{\rm ex}\approx C_{\rm S}$.
The derived ratios of $\displaystyle \frac{P_{\rm ram}}{P_{\rm th}}$ 
and $\displaystyle \frac{P_{\rm rad}}{P_{\rm th}}$
appear to be consistent. 
Both the thermal and radiation pressures  
play a critical role in the expansion of the ionized gas 
and may accelerate ionized outflow. However both the expansion 
and the plausible HII outflow are thought to be confined
in the dense ambient medium \citep{dpre98}.  If the UC HII region is 
in pressure equilibirium with the ambient medium \citep{xie96}, 
we have,
\begin{eqnarray} 
\displaystyle 2kn_{\rm e}T_{\rm e}\xi=n_{\rm H}\left( \mu m_{\rm H}\sigma_V^2 +kT_{\rm k}\right),
\end{eqnarray}
where $\sigma_{\rm V}$ is the turbulent velocity, $T_{\rm k}$ is the kinetic 
temperature of the ambient gas, approximately equal to
the dust temperature $T_{\rm d}$, and $\displaystyle \xi=1+\frac{P_{\rm ram}}{P_{\rm th}}$ 
considering the contribution from ram pressure. From Equation (17),  
we can find a minimum density $n_{\rm H}$  required to confine 
the HII gas,
\begin{eqnarray}
n_{\rm H} &>& 4\times10^8 \xi \left(n_{\rm e}\over 10^6~{\rm cm^{-3}}\right)
\left(T_{\rm e}\over10^4~K\right)\, \nonumber \\
            & &\left(\Delta V_{\rm FWHM}\over
{\rm km~s^{-1}}\right)^{-2}~~~~ {\rm cm^{-3}}
\end{eqnarray}   
where $\displaystyle \Delta V_{\rm FWHM}=\sqrt{(8{\rm ln}2)
\left(\sigma_{\rm V}^2+\frac{kT_{\rm K}}{\mu m_{\rm H}}\right)}$,
the FWHM line width of a molecular line from the surrounding medium.
For $T_{\rm K}\approx\,T_{\rm d}=500~K$, the thermal width is about 
3 km s$^{-1}$. From the SMA measurements of the FWHM line widths 
for the sixteen kinematic absorption features in 
$\rm {H_2CO(3_{30}-2_{02})}$
and $\rm {H_2CO(3_{31}-2_{20})}$  
towards the F1-4 complex \citep{qin08}, 
we find a variance-weighted
mean  $\Delta V_{\rm FWHM}=5\pm0.2$ km s$^{-1}$, 
suggesting a dominant turbulent motion in the line broadening.
Thus, from Equation (18) and the parameters derived 
for the HII regions (Table 3), a minimum 
$n_{\rm H}\approx 2\times10^7$ cm$^{-3}$ is needed to 
confine F3 (component B), which is a factor of a few 
to 10 less than the densities inferred for 
the protostellar cores but is in good agreement with 
the value of $n_{\rm H}$ derived from 
the isothermal model  using the observed $T_{\rm b}= 14$ K 
of the lowest contour in Figure \ref{fig:kin}. 
Our results agree with the predicted range of the molecular 
$n_{\rm H}$  between $10^7-10^8$ cm$^{-3}$  \citep{dpre98}.
\vskip 5pt
\noindent {\bf F1 (H26$\alpha$-3,  H26$\alpha$-4, and H26$\alpha$-6)}
shows complicated kinematics. The line flux from each of 
the three H26$\alpha$ sources suggests the presence of at 
least three O type stars in this region. From analysis of 
the high-resolution VLA image, \cite{dpre98} suggested a 
total of seven late O-type or early B-type stars in F1. 
The observed kinematics suggest that complicated dynamics 
of ioinzed gas are involved in the OB association.      
\vskip 5pt
\noindent {\bf G (H26$\alpha$-8)} also shows a broad wing 
in its H26$\alpha$ line spectrum, which corresponds to a
large radial velocity gradient ($\Delta V_{\rm r}$)  northeast-to-southwest 
across the source with a size  0.6\arcsec\, as seen in 
the moment 1 image (see Figure \ref{fig:kin}).
Taking the parameters inferred from our modeling (Table 3) and $L_*$(O8.5)$\approx5.4\times10^4\,L_\sun$, 
with Equation (15), we find that 
${\displaystyle \frac{P_{\rm rad}}{P_{\rm th}}\approx1}$  and $\approx0.6$
for the hyper-compact component (A) and the larger component (B), 
respectively. The observed velocity gradient in the component B region
appears to be mainly due to a bipolar ionized outflow accelerated 
in the region surrounding an O8.5 star. The structure of kinematics 
for the hyper-compact component (A) with a weak stimulated source as suggested 
earlier  was not resolved in these observations.    
\vskip 5pt
\noindent {\bf F10.37 (H26$\alpha$-5)} shows a broad H26$\alpha$ 
line profile with $\Delta V_{\rm FWHM}=51\pm6$ km s$^{-1}$
but no obvious radial velocity gradient is seen in the moment 1 
image (see Figure \ref{fig:kin}). For $T_e=1\times10^4$ K, the 
thermal broadening gives $\Delta V_{\rm FWHM}^{\rm th}=$21 km s$^{-1}$.
Subtracting the thermal broadening, the residual line width
$\Delta V_{\rm FWHM}^{\rm nth}=$46 km s$^{-1}$ requires large 
nonthermal motions. A bipolar ionized outflow along the 
line-of-sight might be responsible for the large velocity dispersion 
$\sigma_{\rm V} = \frac{\displaystyle 1}{\displaystyle \sqrt{8{\rm ln}2}}\Delta V_{\rm FWHM}^{\rm nth}=20$
km s$^{-1}$ in this region.  

 % -----------edits to here 14 apr  18:40 PST    mchw. on MacBook Pro ---------

\subsection{Formation of massive protostellar cores}
In addition to the H26$\alpha$ sources, there are about two dozen
protostellar cores with  masses much greater than the corresponding 
Bonnor-Ebert masses, and  a total molecular mass in the range
5 -- 223 $M_\odot$. The observed filamentary structures in the 
continuum emission at the submillimeter wavelength suggest that 
the massive stars and protostellar cores were formed through 
fragamention during the initial collapse of the massive core in 
Sgr B2 Main \citep{qin08}. The inferred surface density of each 
protostellar core appears  to be greater than unity;
a minimum mass for forming a massive star in order to avoid 
further fragamentation \citep{krum08}. Through  competitive 
accretion, each of the protostellar cores will grow by adding
an accreted mass  ${\dot M}\, t_{ff}$ within the free-fall timescale
$t_{ff}$. The accretion rate can be estimated for an isothermal 
flow in which the density, $\rho(r)\propto\,r^{-3/2}$, and the 
enclosed mass within a radius  $r$, $M(r)\propto\,r^{3/2}$,  
can be described by a power law \citep{shu77}. The accretion 
rate is linearly proportional to the accretion velocity $V_{\rm Acc}$ 
and the enclosed mass $M_{\rm R}$ within $r=R$ but inversely 
proportional to R,
\begin{eqnarray}
\displaystyle {\dot M}&=&V_{\rm Acc}\frac{dM(r)}{dr}\Bigr|_{r=R} \, \nonumber \\
                      &=&0.3 M_\sun\,y^{-1}\left(V_{\rm Acc}\over {\rm km\,s^{-1}}\right)\left(M_{\rm R}\over {M_\sun}\right)
                         \left(R\over  {\rm AU}\right)^{-1}.
\end{eqnarray}
Based on our measurements  
$2R\approx\sqrt{5\arcsec\times3\arcsec}\approx3\times10^4$ AU,
$M_{\rm R}\sim1.2\times10^3\, M_\odot$, and 
$V_{\rm Acc}\sim C_{\rm S}=\sqrt{\displaystyle {kT_{\rm K}}\over{
\displaystyle  \mu m_{\rm H}}}\approx1.3$ km s$^{-3}$ 
for a kinetic temperature  $T_{\rm K}\approx{T}_{\rm d}=500$ K,   
we inferred ${\dot M}\approx$0.03 $M_\odot\,y^{-1}$.  
For a constant accretion rate of 0.03 $M_\odot\,y^{-1}$, 
in a mean free-fall time-scale  $\overline{t}_{ff}\approx 2\times10^3$ y,
about 60 $M_\odot$, or $\sim 5$\% of the total mass increases in 
the protostellar cores. 

 % -----------edits to here 14 apr  21:16 PST    mchw. on MacBook Pro ---------

We have observed a snapshot of the process of forming a 
massive-star cluster. A few points regarding the formation of 
massive star cores can be drawn from our analysis of
the SMA observations. First, Figure \ref{fig:kin} shows a 
noticeable angular offset of $\sim0.1$\arcsec\, between the 
H26$\alpha$ sources (H26$\alpha$-1,  H26$\alpha$-4 and  
H26$\alpha$-6) from their continuum counterparts (Smm-1, 
Smm-4 and Smm-10). The relative angular offsets are 
significant ($>5\sigma$), 
corresponding to a projected distance of 800 AU. If the 
H26$\alpha$ sources denote the locations of the new born O stars,
the offsets of the protostellar cores (Smm-1, Smm-4 and Smm-10) 
from the ionizing stars might suggest that the local accretion 
centers are altered after the onset of the radiation from the O stars. 
A massive protostellar core in the immediate vicinity of 
a UC HII region has been also observed in the W51e2 complex 
\citep{SHI10a,SHI10b}. The authors use SMA high-resolution 
observations to show that the protostellar core (W51e2-E) with a mass of 
140 $M_\odot$,  which is located 0.9\arcsec\, 
(4600 AU) from the UC HII region (W51e2-W),  becomes the new accretion center and 
undergoes active star formation ejecting a powerful 
molecular outflow. Multiple pairs of O-type star 
and massive protostellar core found in Sgr B2 Main 
show a cluster version of the W51e2 system. 
Consequently,  a considerable amount of mass  must have 
existing in each of the paired 
protostellar cores before the 
O-type stars were born unless there were large accretion rates,
$\sim$0.1 $M_\odot$ s$^{-1}$ at each of the massive protostellar cores 
in the past 1$\times10^3$ years after the onset of  radiation from the O stars. 
The protostellar cores and their companion  
massive stars likely originate  from 
fragmentation during the initial collapse of the cloud. 
After the primitive molecular clumps formed, 
the massive ones undergo  substantial gravitational collapse to form stars. 
Competitive accretion may occur in 
the process of forming a cluster of massive stars 
in a massive core like Sgr B2 Main. However, from our observations
of this particular phase of massive star formation in a cluster,
it is uncertain how much mass has been added to each of 
the protostellar cores through competitive accretion
owing to the uncertainty in the time-scale of the process.
On the basis of our analysis, accretion will add a small
fraction of mass to each of the protostellar cores in a 
mean free-fall time-scale of 2$\times10^3$ y.

\section{Conclusion and Summary} 

Using SMA archival data observed in 2007 for Sgr B2 Main,
we imaged the H26$\alpha$ line and continuum emission at 0.86 mm 
with natural and uniform weighted synthesized beams of
$0.40\arcsec\times0.28\arcsec$ and $0.36\arcsec\times0.22\arcsec$,
respectively. Eight compact H26$\alpha$ line sources have been detected
towards the central 5\arcsec\, ($\sim0.2$ pc). Using the non-LTE approach, 
we fitted the H26$\alpha$ line along with other available RRL and 
radio continuum data with two isothermal, homogeneous HII components:
1) high-density (a few $\times10^7$ cm$^{-3}$), hyper-compact 
(several hundred AU) and 2) lower-density (a few $\times10^6$ cm$^{-3}$), 
ultra-compact (a few thousand AU). The observed H26$\alpha$ line 
fluxes suggest that each of the eight H26$\alpha$ line
emission regions must be powered by a Lyman continuum source 
from an O type star or a cluster of B-type stars. The typical 
age of the HII regions are a thousand years. The brightest 
H26$\alpha$-1 source in F3 appears to be associated with an O6 star. 
The H26$\alpha$ line from H26$\alpha$-7 (F10.39) appears to be 
enhanced by stimulated emission in the vicinity of an O9.5 star.

About two dozen compact continuum emission cores are detected 
at 0.86 mm in the central 5\arcsec\, (0.2pc). Using a power-law 
distribution of dust temperature and observed brightness 
temperature, we determined the physical properties for each of 
the protostellar cores with masses in the range of 5 to $\gtrsim$ 200 
solar masses. The surface densities of the prostellar cores are in the range 
13 to 150 $g\, {\rm cm^{-2}}$, 
and molecular densities 1$\times10^8$ to 1$\times10^9$ ${\rm cm^{-3}}$.
We also calculated the Bonnor-Ebert mass ($M_{\rm BE}$)  for each core. The total 
molecular mass of each core appears to be much greater than the value 
of $M_{\rm BE}$, suggesting that the cores undergo a  substantial
gravitational collapse to form massive stars
with a typical free-fall time scale of a few thousand years.   
%%%%%%%%%%%%%%%%%%%%%%%%%%%%%%%%%%%%%%%%%%%
%%%%%%%%%%%%%%%%%%%%%%%%%%%%%%%%%%%%%%%%%%%
\acknowledgments
We are grateful to the anonymous referee for thoughtful comments
and suggestions which were helpful for improving this paper.
We thank Dr. Hui Shi for her assistance preparing Miriad scripts 
for the reduction of the SMA data. JHZ is grateful to the National 
Astronomical Observatories of China for hosting his visit during 
the course of writing this research paper.

\appendix
\section{Lyman continuum photon flux}
In this Appendix, we show the detailed procedure to
derive the Lyman continuum photon flux from the measurements
of a hydrogen radio recombination line.
\subsection{Photoionizations and recombinations of hydrogen in HII regions}
For an HII region with electron density ${\rm n_e}$, 
the ionization equilibrium can be described by the balance 
between photoionizations and recombinations of electrons with 
the ions. The  number of ionizing photons ($N_{\rm I}$) 
emitted by a star in unit time, or the Lyman continuum photon
flux
($N_{\rm Lym}$), is equal to the number of recombinations 
($N_{\rm R}={\displaystyle 4\pi\over3}R_{\rm s}^3 {\rm n_e}^2 \alpha_{\rm B}$)
to the excited levels within the Str{\"o}mgren radius ($R_{\rm s}$) 
in unit time \citep{oste74}:

\begin{eqnarray}
\displaystyle {N_{\rm Lym}}&=&
                         {\displaystyle 4\pi\over3}R_{\rm s}^3 n_e^2 \alpha_{\rm B}
\end{eqnarray}
where $\alpha_{\rm B}=\alpha_{\rm B}(T_{\rm e})$ is the recombination coefficient in case B,
which is a
function of electron temperature ($T_{\rm e}$) with a value 
$\alpha_{\rm B}=2.58\times10^{-13}$ cm$^{-3}$ s$^{-1}$ 
at $T_{\rm e}=1\times10^4$ K \citep{humm63};
and $R_{\rm s}^3 \approx 
{\displaystyle 3\over{\displaystyle 4\pi} }V\approx {\displaystyle 3\over
{\displaystyle 4\pi}} \Omega_{\rm HII} D^2 Lf_{\rm V}$.

For an HII region located at distance $D$, with solid angle
 $\Omega_{\rm HII}$, path length $L$, and electron volume filling factor $f_{\rm V}$,
 Equation (A1) can be rewritten as,

\begin{eqnarray}
\displaystyle {N_{\rm Lym}}&=&
                       n_e^2Lf_{\rm V} \alpha_B \Omega_{\rm HII} D^2
\end{eqnarray} 

\subsection{Optical depth and flux density of hydrogen radio recombination lines}

For an isothermal-homogeneous HII region, the optical depths
for  hydrogen radio recombination lines
($\tau_{\rm L}$) and free-free continuum ($\tau_{\rm C}$), 
are proportional to the emission measure  ($EM=n_{\rm e}^2 L f_{\rm V}$),
and are given by \cite{shaver75}:
\begin{eqnarray}
\displaystyle
\tau_L & =     & b_{\rm N}\beta_{\rm N}\tau_{\rm L}^* 
       \approx 575 b_{\rm N}\beta_{\rm N} \left(\nu\over {\rm~GHz}\right)^{-1}
\left(n_e\over {\rm~cm}^{-3}\right)^2  
       \left(Lf_V \over {\rm pc}\right)\left(T_e\over {\rm K}\right)^{-5/2}
\left(\Delta V_{{\rm D}}\over {\rm km~s}^{-1}\right)^{-1} 
       \left(1+1.48{{\Delta V_P}\over{\Delta V_D}}\right)^{-1}~~,
\end{eqnarray}

\noindent and

\begin{eqnarray}
\displaystyle
\tau_C &\approx& 0.08235 \left(n_e\over {\rm~cm}^{-3}\right)^2 
\left(Lf_V\over {\rm pc}\right)  
\left(\nu\over {\rm~GHz}\right)^{-2.1} 
       \left(T_e\over {\rm K}\right)^{-1.35} a(\nu,T_e)~~,
\end{eqnarray}

\noindent where $\tau_{\rm L}^*$ is the LTE line optical depth; 
$b_{\rm N}$ and $\beta_{\rm N}$ are the population departure coefficients;
$n_e$ is the electron density; $\Delta V_{\rm D}$ and $\Delta V_{\rm P}$ are the
FWHM Doppler and pressure line widths in \kms, respectively. 
The correction factor $a(\nu,T_e)$ is of order unity.
 
On the other hand, 
if an HII region,  with no dust in it,  is  optically thin and LTE ($\tau_{\rm C} << 1$, 
$\tau_{\rm L}^* << 1$, and $b_{\rm N}=\beta_{\rm N}=1$), 
the line flux density 
from Equation (1) in the main text becomes,

\begin{eqnarray}
S_{\rm L}^*&\approx&{{2k\nu^2}\over c^2}\Omega_{\rm HII} T_{\rm e}\tau_{\rm L}^*
\end{eqnarray}

Normally, the Lyman continuum photon flux ($N_{\rm Lym}$) is
derived from the emission measure determined from the free-free continuum
flux density observed at a radio frequency, {\it e.g. } \cite{wilc94}. 
In the following, we show the derivation
of the emission measure from a radio recombination line flux. Therefore,
$N_{\rm Lym}$ can be determined directly from  a radio recombination line flux.   

\subsection{Emission measure from hydrogen radio recombination lines}

From Equations (A3) and (A5), neglecting the pressure broadening ($\Delta V_{\rm D}\gg\Delta V_{\rm P}$)
the emission measure for an optically-thin, LTE gas can be expressed as,

\begin{eqnarray}
n_e^2Lf_{\rm V}
  &=& 5.66\times10^{-8} {\rm cm^{-6}\, pc}
 \left(S_{\rm L}^* \Delta V_{\rm D} \over {\rm Jy\,km\,s^{-1}}\right) 
\left(\nu\over {\rm~GHz}\right)^{-1} \left(T_e\over {\rm K}\right)^{3/2}
 \Omega_{\rm HII}^{-1} 
\end{eqnarray}

For a general isothermal, homogeneous HII region, the emission measure
can be determined by the following equation,

\begin{eqnarray}
n_e^2Lf_{\rm V}
  &=& 5.66\times10^{-8} {\rm cm^{-6}\, pc}
 \left(S_{\rm L} \Delta V_{\rm D} \over {\rm Jy\,km\,s^{-1}}\right) 
 \left(\nu\over {\rm~GHz}\right)^{-1} \left(T_e\over {\rm K}\right)^{3/2}
 \Omega_{\rm HII}^{-1} \Psi_{\rm N}^{-1}
\end{eqnarray}
\noindent where $\Psi_{\rm N}$ is the ratio of the  non-LTE line flux
($S_{\rm L}$) to the LTE line flux in optically thin 
($S_{\rm L}^*[\tau_{\rm L}^*<<1$]) for the line transition 
${\rm N+1 \rightarrow N}$, where N is the principal qauntum number.
$\Psi_{\rm N}$ is a function of the line and continuum optical depths
($\tau_{\rm L}$ and $\tau_{\rm C}$) and departure coefficients 
($b_{\rm N}$ and $\beta_{\rm N}$),

\begin{eqnarray}
\Psi_{\rm N} &\equiv& {S_{\rm L}\over S_{\rm L}^*[\tau_{\rm L}^*<<1]} 
     \approx\Bigg[\left(\tau_{\rm L}/\beta_{\rm N} +\tau_{\rm C} \over 
       \tau_{\rm L} +
\tau_{\rm C} \right)\left(1-e^{-(\tau_{\rm L}+\tau_{\rm C})}\right) 
        -(1- e^{-\tau_{\rm C}})\Bigg]\tau_{\rm L}^{*-1},
\end{eqnarray}
neglecting the term associated with 
the dust emission ($S_{\rm d}$) in Equation (1) of the main text.

\subsection{Lyman continuum photon flux from hydrogen radio recombination lines}

For an electron temperature ($T_{\rm e}$) in an isothermal, homogeneous 
HII region, inserting $n_{\rm e}^2Lf_{\rm V}$ of 
Equation (A7) into Equation (A2), we derived
the Lyman continuum photon flux from the measurements with
flux density ($S_{\rm L}$) and FWHM of Doppler brodening
($\Delta V_{\rm D}$) for a hydrogen radio recombination
line (${\rm N+1 \rightarrow N}$) at frequency $\nu_{\rm L}$,
\begin{eqnarray}
N_{\rm Lym}
 &=& 4.29\times10^{47}\left({S_{\rm L}\Delta V_{\rm D}}\over{\rm Jy~km~s^{-1}}\right)
\left({\nu_{\rm L}\over {\rm GHz}}\right)^{-1}\left({D}\over{\rm kpc}\right)^2 
            \left({T_{\rm e}}\over{10^4\,K}\right)^{1.5}
       \left(\alpha_{\rm B}(T_{\rm e}) \over\alpha_{\rm B}(10^4\,K)\right)
     \Psi_{\rm N}^{-1} ~~~~ph~s^{-1} \,
\end{eqnarray}
where $\Psi_{\rm N}$ corrects for the effects due to the optical depths and 
non-LTE distribution.
Thus, from Equation (A8), $\Psi_{\rm N}\approx1$ for optically thin, LTE gas;
$\Psi_{\rm N}\approx b_{\rm N}(1-{1\over2}\tau_{\rm C}\beta_{\rm N})$ for optically 
thin,non-LTE gas; and $\displaystyle \Psi_{\rm N}\approx {e^{-\tau_{\rm C}}(1-e^{-\tau_{\rm L}^*})\over \tau_{\rm L}^*}$
for the LTE gas becomes optically thick.

% Edited appendix text  June 07, 12:32, but did not check equations.  MCHW

% -----------edits to here 18 apr  11:16 PST    mchw. on MacBook Pro ---------
\vskip 10pt
%

%\onecolumn
\clearpage
%\documentclass[10pt,preprint]{aastex}
%\begin{document}
\begin{deluxetable}{lcc}
\tablenum{1}
\tablecolumns{3}
\tabletypesize{\scriptsize}
\tablecaption{Log of observations and imaging}
\tablewidth{0pc}
\tablehead{ 
\\
\multicolumn{1}{l}{Parameters}&
\multicolumn{1}{c}{H$26\alpha$}&
\multicolumn{1}{c}{Continuum} \\
} 
\startdata
\multicolumn{3}{l}{OBSERVATION:}         \\
Date                  & \multicolumn{2}{c}{2007 June 18}\\
Array configuration   & \multicolumn{2}{c}{Very extended (8 antennas)}\\
Pointing R.A.  (J2000)& \multicolumn{2}{c}{\,\,\,\,\,17:47:20.151}\\ 
Pointing decl. (J2000)& \multicolumn{2}{c}{--28:23:04.79}\\
$\nu_{\rm LO}$        & \multicolumn{2}{c}{348.664 GHz}\\
$\delta\nu$           & \multicolumn{2}{c}{0.406 MHz}\\
Bandwidth             & \multicolumn{2}{c}{2GHz(USB) \& 2GHz(LSB)}\\
On-source time        & \multicolumn{2}{c}{1.95 hr}\\
$T_{\rm sys}$         & \multicolumn{2}{c}{150-500 K}\\
\\
\multicolumn{3}{l}{CALIBRATORS:} \\
 Flux density         & \multicolumn{2}{c}{Callisto}\\
 Phase                & \multicolumn{2}{c}{J1733-130}\\
 Bandpass             & \multicolumn{2}{c}{J1229+020, J1733-130, J1743-038, J1751+096, J2015+371} \\
\\
\multicolumn{3}{l}{IMAGING:}       \\
{$\nu_0$\,\,\,\.  or\,\,\,\, $\lambda$}& 353.623 GHz& 0.86 mm  \\
channel width         & 3\,\,km\,s$^{-1}$  & \dots \\
FWHM beam             & 0.40\arcsec$\times$0.28\arcsec (15\arcdeg)
                      & 0.36\arcsec$\times$0.22\arcsec (13\arcdeg) \\
r.m.s.                & 0.1 Jy bm$^{-1}$ chan$^{-1}$ &
                         8  mJy bm$^{-1}$    \\
\enddata
\end{deluxetable}
%\end{document}

\clearpage
%\documentclass[10pt,preprint]{aastex}
%\begin{document}
\begin{deluxetable}{lcrccccc}
\tablenum{2}
\tabletypesize{\scriptsize}
\tablecaption{Submillimeter Properties of Sgr B2 (Main)}
\tablewidth{0pt}
\tablehead{
\multicolumn{8}{c}{The H26$\alpha$ Line Measurements}\\
\cline{4-5} \\
\colhead{Region}&
\colhead{$ {\Delta\alpha}$\tablenotemark{a}} &
\colhead{$ {\Delta\delta}$\tablenotemark{a}} &
\colhead{$\theta_{Maj}\times\theta_{Min}, PA$}&
\colhead{$ {S_{\rm H26\alpha}}$}&
\colhead{$V_{\rm LSR}$}&
\colhead{$\Delta V_{\rm FWHM}$}&
\colhead{$S_{\rm C}$} \\
\colhead{}&
\colhead{($\arcsec$)}&
\colhead{($\arcsec$)}&
\colhead{($\arcsec\times\arcsec$,\,\,$\arcdeg$)}&
\colhead{(Jy)}&
\colhead{(km s$^{-1}$)}&
\colhead{(km s$^{-1}$)}&
\colhead{(Jy)} } 
\startdata
%\tableline \\
H26$\alpha$-1 (F3) &--0.17&+0.03&0.29$\times$0.21,\,\,\,--25&7.0\,\,\,$\pm$0.1\,\,\,&
64$\pm$1&40$\pm$\,\,\,1&2.9$\pm$0.5 \\
H26$\alpha$-2 (F4)& +0.37&+0.37&U&0.63$\pm$0.08&66$\pm$2&35$\pm$8&
0.41$\pm$0.10\\
H26$\alpha$-3 (F1A)&--0.53&+0.73&U&1.2\,\,\,$\pm$0.1\,\,&48$\pm$2&54$
\pm$5&1.8$\pm$0.4 \\
H26$\alpha$-4 (F1B)&--0.92&+0.38&U&0.8\,\,\,$\pm$0.1\,\,&73$\pm$1&36$
\pm$4&1.0$\pm$0.4 \\
H26$\alpha$-5 (F10.37)&--0.00&--1.18&U&0.62$\pm$0.08&52$\pm$2&51$\pm$6
&0.77$\pm$0.10\\
H26$\alpha$-6 (F10.303)&--0.90&+1.08&U&0.74\,\,\,$\pm$0.1\,\,&57$\pm$2
&50$\pm$6&0.91$\pm$0.24 \\
H26$\alpha$-7 (F10.39)&+0.22&--1.97&U&0.62$\pm$0.09&69$\pm$2&29$\pm$6
& 0.35$\pm$0.09\\
H26$\alpha$-8 (G)     &+1.45&+1.70&U&0.73\,\,\,$\pm$0.1\,\,&53$\pm$2&
48$\pm$6&0.2$\pm$0.1\\
\\
\tableline \\
\multicolumn{8}{c}{The Continuum Measurements at 0.8 mm}\\
\cline{3-6} \\
\colhead{Source}&
\colhead{$ {\Delta\alpha}$\tablenotemark{a}} &
\colhead{$ {\Delta\delta}$\tablenotemark{a}} &
\colhead{$\theta_{Maj}\times\theta_{Min}, PA$}&
\colhead{$S_{\rm P}$}&
\colhead{$S_{\rm T}$}&
\colhead{$ {\Delta \theta}$\tablenotemark{a}}&
\colhead{$T_{\rm b}$} \\
\colhead{}           &
\colhead{($\arcsec$)}&
\colhead{($\arcsec$)}&
\colhead{($\arcsec\times\arcsec$,\,\,$\arcdeg$)}&
\colhead{(Jy beam$^{-1}$)}&
\colhead{(Jy)} &
\colhead{($\arcsec$)}&
\colhead{($10^2$ K)}
\\
Smm-1& --0.20  & +0.16 &0.80$\times$0.40,  +10 &2.6$\pm$0.2&  11$\pm$0.8&  0.26& 3.4  \\
Smm-2&  +0.78  & +0.09 &0.22                   &0.4$\pm$0.1& 0.4$\pm$0.1&  0.79& 0.8  \\
Smm-3& --0.39  &--0.72 &0.35$\times$0.25,  +3  &2.2$\pm$0.2& 2.2$\pm$0.2&  0.82& 2.5  \\
Smm-4& --0.49  & +0.68 &0.83$\times$0.30,  +32 &2.0$\pm$0.3& 7.0$\pm$0.9&  0.84& 2.7  \\
Smm-5&  +1.00  & +0.41 &0.22                   &0.4$\pm$0.1& 0.4$\pm$0.1&  1.08& 0.8  \\
Smm-6& --0.92  & +0.59 &0.35$\times$0.26,  +21 &1.1$\pm$0.2& 2.2$\pm$0.3&  1.09& 2.4  \\
Smm-7& +0.01  &--1.20  &0.61$\times$0.31, --31 &0.9$\pm$0.1& 2.7$\pm$0.3&  1.20& 1.4  \\
Smm-8& --0.68  &--1.09 &0.22                   &0.8$\pm$0.2& 0.8$\pm$0.2&  1.28& 1.6 \\
Smm-9&  +0.58  &--1.17 &0.30$\times$0.15, --75 &0.5$\pm$0.1& 0.6$\pm$0.1&  1.31& 2.0 \\
Smm-10&--1.00  & +1.01 &0.38$\times$0.22, --21 &1.3$\pm$0.2& 1.7$\pm$0.2& 1.42&  2.0  \\
Smm-11&  +1.03  & +1.04&0.22                   &0.3$\pm$0.1& 0.3$\pm$0.1& 1.46&  0.6 \\
Smm-12& --0.08  & +1.59&0.53$\times$0.14, --82 &0.4$\pm$0.1& 0.8$\pm$0.2& 1.59&  1.1 \\
Smm-13& --0.77  &--1.57&0.53$\times$0.45, --75 &1.2$\pm$0.2& 3.9$\pm$0.6& 1.75&  1.6  \\
Smm-14&  +0.33  &--1.74 &0.39$\times$0.19, --48&0.8$\pm$0.2& 1.3$\pm$0.3& 1.77&  1.7 \\
Smm-15& --1.77  &--0.02 &0.45$\times$0.09, +29 &0.4$\pm$0.2& 0.4$\pm$0.2& 1.77&  1.0 \\
Smm-16&  +0.83  &--1.67 &0.22                  &0.4$\pm$0.1& 0.4$\pm$0.1& 1.86&  0.8 \\
Smm-17& --0.89  & +1.64 &0.53$\times$0.20, --13&0.8$\pm$0.1& 1.6$\pm$0.2& 1.87&  1.5 \\
Smm-18& --1.77  & +1.24 &0.22                  &0.4$\pm$0.1& 0.4$\pm$0.1& 2.16&  0.8 \\
Smm-19&  +0.53  &--2.12 &0.22                  &0.7$\pm$0.1& 0.7$\pm$0.1& 2.19&  1.4 \\
Smm-20& --2.12  & +0.69 &0.22                  &0.4$\pm$0.1& 0.4$\pm$0.1& 2.23&  0.8 \\
Smm-21&  +1.46  & +1.74 &0.22                  &0.2$\pm$0.1& 0.2$\pm$0.1& 2.27&  0.4 \\
Smm-22& --1.57  & +2.04 &0.30$\times$0.19, +27 &0.5$\pm$0.2& 0.5$\pm$0.2& 2.57&  0.9 \\
Smm-23&  +0.56  &--2.57 &0.60$\times$0.22, +57 &0.4$\pm$0.1& 1.0$\pm$0.1& 2.63&  0.7 \\  
\enddata
\tablenotetext{a}{Offsets from F3 at RA(J2000)=17:47:20.178, Decl(J2000)=--28:23:04.67}
\end{deluxetable}
%\end{document}

\clearpage
\includegraphics{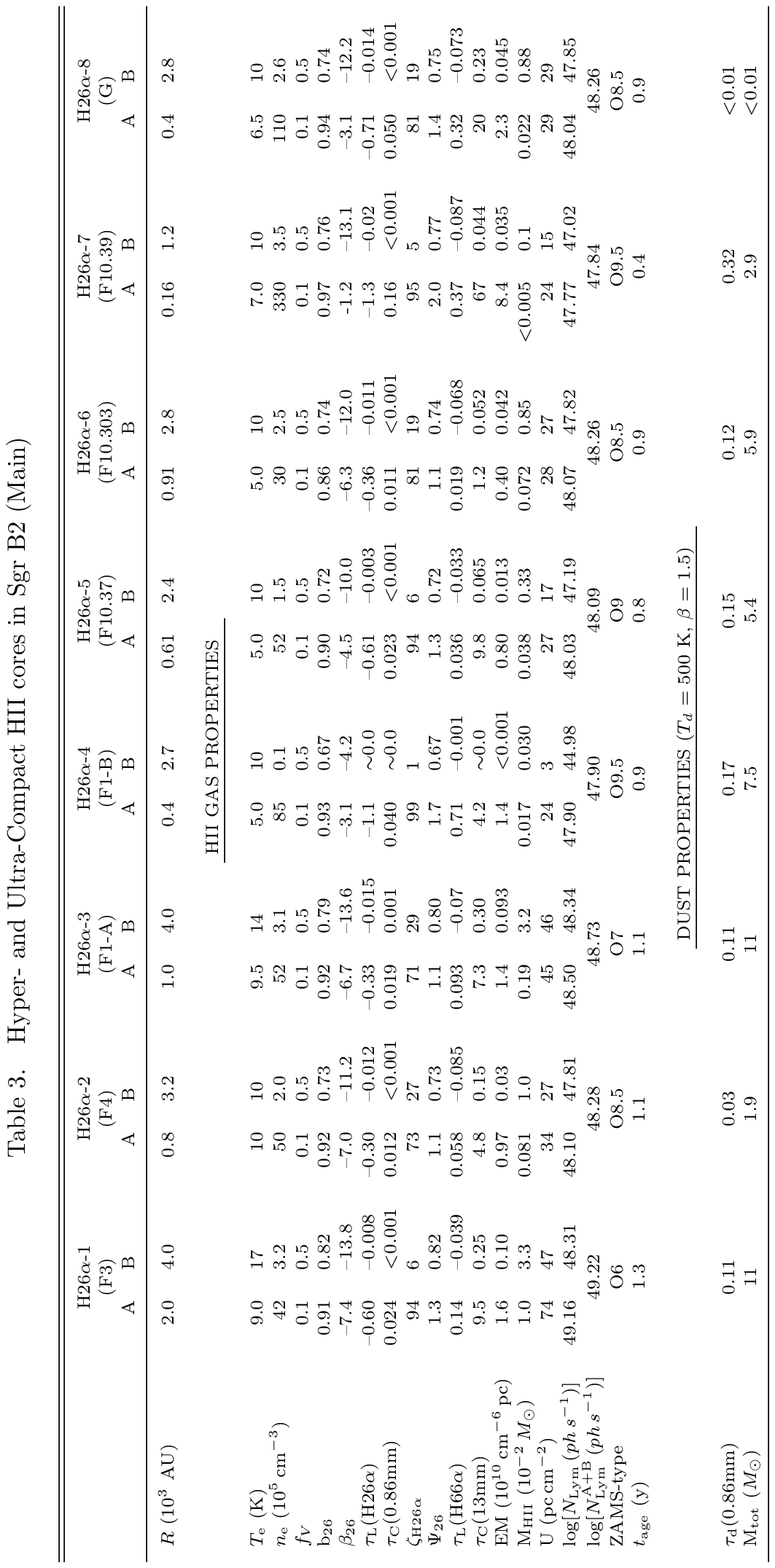}
\clearpage
%\documentclass[10pt,preprint]{aastex}
%\begin{document}
\begin{deluxetable}{lcccrrrccc}
\tablenum{4}
\tabletypesize{\scriptsize}
\tablecaption{Properties of protostellar cores in Sgr B2 (Main)}
\tablewidth{0pt}
\tablehead{
\colhead{Name}&
\colhead{r} &
\colhead{size} &
\colhead{$T_{\rm d}$}&
\colhead{$\tau_{\rm d}$}&
\colhead{$M$}&
\colhead{$\Sigma$}&
\colhead{$n({\rm H})$}&
\colhead{$M_{\rm BE}$}&
\colhead{$t_{ff}$}
\\
\colhead{}           &
\colhead{(pc)}       &
\colhead{($10^3$ AU)}&
\colhead{($10^2$ K)}&
\colhead{}&
\colhead{($M_\odot$)}&
\colhead{($g~{\rm cm^{-2}}$)}&
\colhead{($10^8 \,{\rm cm^{-3}}$)}&
\colhead{($M_\odot$)}&
\colhead{($10^3$ yr)} 
} 
\startdata
 Smm-1 &0.01&4.5&  5.0&  1.1& 143&  55&  2.1&  2.9&  2.3\\
 Smm-2 &0.05&1.8&  2.5&  0.4&   8&  20&  1.9&  1.0&  2.4\\
 Smm-3 &0.04&2.4&  2.6&  2.6&  93& 130&  9.5&  0.5&  1.1\\
 Smm-4 &0.04&4.4&  2.7&  1.9& 223&  92&  3.7&  0.9&  1.8\\
 Smm-5 &0.07&1.8&  2.2&  0.4&   9&  22&  2.2&  0.9&  2.3\\
 Smm-6 &0.05&2.4&  2.5&  3.0& 111& 150& 10.8&  0.4&  1.0\\
 Smm-7 &0.07&3.5&  2.2&  1.0&  74&  48&  2.4&  0.8&  2.2\\
 Smm-8 &0.07&1.8&  2.2&  1.3&  25&  63&  6.2&  0.5&  1.4\\
 Smm-9 &0.08&1.7&  2.1&  0.9&  17&  46&  4.8&  0.5&  1.6\\
 Smm-10&0.07&2.3&  2.2&  2.3&  77& 112&  8.4&  0.4&  1.2\\
 Smm-11&0.09&1.8&  2.1&  0.3&   7&  17&  1.7&  0.9&  2.6\\
 Smm-12&0.09&2.2&  2.1&  0.7&  21&  35&  2.8&  0.7&  2.0\\
 Smm-13&0.09&3.9&  2.0&  1.6& 155&  79&  3.5&  0.6&  1.8\\
 Smm-14&0.10&2.2&  1.9&  2.1&  64& 105&  8.4&  0.4&  1.2\\
 Smm-15&0.09&1.6&  2.0&  0.6&  10&  32&  3.4&  0.6&  1.8\\
 Smm-16&0.11&1.8&  1.9&  0.6&  11&  27&  2.7&  0.6&  2.1\\
 Smm-17&0.10&2.6&  2.0&  1.4&  59&  68&  4.5&  0.5&  1.6\\
 Smm-18&0.11&1.8&  1.9&  0.6&  11&  28&  2.7&  0.6&  2.1\\
 Smm-19&0.12&1.8&  1.8&  1.5&  30&  75&  7.4&  0.3&  1.2\\
 Smm-20&0.11&1.8&  1.9&  0.6&  11&  28&  2.8&  0.6&  2.0\\
 Smm-21&0.13&1.8&  1.8&  0.3&   5&  13&  1.3&  0.8&  3.0\\
 Smm-22&0.13&1.9&  1.8&  0.7&  15&  33&  3.0&  0.5&  2.0\\
 Smm-23&0.15&2.9&  1.7&  0.6&  31&  28&  1.7&  0.6&  2.6\\
\enddata
\end{deluxetable}
%\end{document}

\end{document}